\documentclass[preprint]{aastex631}

\submitjournal{ApJ}

\shorttitle{Sheared/guide field asymmetric reconnection}
\shortauthors{Nitta \& Kondoh}
\graphicspath{{./}{figs/}}

\begin{document}

\title{Effects of magnetic shear and thermodynamic asymmetry on spontaneous MHD magnetic reconnection}

\author{Shin-ya Nitta}
\affiliation{Solar Science Observatory, 
National Astronomical Observatory of Japan, 
2-21-1 Osawa, Mitaka, Tokyo, 181-8588, Japan}
\affiliation{Tsukuba University of Technology, 4-3-15 Amakubo, Tsukuba, 305-8520, Japan}
\affiliation{nstitute of Space and Astronautical Science, Japan Aerospace Exploration Agency, 3-1-1 Yoshinodai, Chuo-ku, Sagamihara, Kanagawa, 252-5210, Japan}
\email{nittasn@yahoo.co.jp}

\author{Koji Kondoh}
\affiliation{Research Center for Space and Cosmic Evolution, Ehime University, 2-5 Bunkyo-cho, Matsuyama, Ehime, 790-8577, Japan}
\email{kondo@cosmos.ehime-u.ac.jp}

\begin{abstract}

The spontaneous evolution of magnetic reconnection in generalized situations (with thermodynamic asymmetry regarding the current sheet and magnetic shear) is investigated using two-dimensional magnetohydrodynamic simulation. We focus on the asymptotic state of temporal evolution, i.e., the self-similarly expanding phase (Nitta et al. 2001). 1) A long fast-mode shock is generated in front of the shorter plasmoid as in the shear-less thermodynamically asymmetric case; however, the sheared magnetic component weakens the shock. This fast shock may work as a particle acceleration site. 2) The shorter plasmoid-side plasma infiltrates the longer plasmoid across the current sheet. Then, the plasmas from both sides of the current sheet coexist on the same magnetic field lines in the longer plasmoid. This may result in efficient plasma mixing. 3) The thermodynamic asymmetry and magnetic shear drastically decrease the reconnection rate in many orders of magnitude. 

\end{abstract}


\section{Introduction} \label{sec:intro}
Magnetic reconnection (MRX) is widely accepted as a powerful engine process of a magnetized plasma system in a lot of classes of astrophysical phenomena. It is believed that magnetic storms in the geo/planetary magnetospheres (e.g., Russell and McPherron 1973), flares in the solar/stellar atmosphere (e.g., Parker 1963), and intermittent flares in the accretion disk system (e.g., Hayashi, Shibata, Matsumoto 1996) are driven by MRX. 

MRX drastically changes the topological structure of magnetic field lines in a short time. Based on the recent progress in solar physics, it has been elucidated that small-scale successive MRX (so-called tether-cut reconnection) is a key precursor process that induces large-scale flare in strongly twisted magnetic structures in the solar corona (e.g., Moore et al. 2001). The tether-cut reconnection successively and gradually changes the magnetic structure that tethers the huge twisted magnetic fluxes, which possess an enormous magnetic free energy. Finally, the twisted structure reaches a critical marginally stable state and precipitates into a catastrophic macroscopic instability to release an enormous amount of magnetic energy. Thus, the relative speed of the magnetic topology change and magnetic energy release with respect to the speed of other processes is important to determine the manner by which the event would proceed. We have to correctly evaluate the proceeding speed of MRX considering various circumstances.

MRX is also considered as a fundamental process in the phenomenological sense. We do not have a widely accepted standard model of the fast MRX composed of a single reconnection except for the Petschek model (Petschek 1964). Recent trend of MRX studies mainly focus on complicated models including multiple MRX systems (e.g., the turbulent MRX (Lazarian \& Vishniac 1999), plasmoid mediated MRX (Loureiro \& Uzdensky 2016), sporadic Petschek-type MRX (Shibayama et al. 2015, 2019), and plasmoid induced MRX/fractal MRX (Shibata 1996 and Shibata \& Tanuma 2001)); however, these are out of the scope of this study. We focus on the MRX as an elementary process of plasma astrophysics, thus, we concentrate on the single MRX system. We discuss this point later in Section \ref{sec:rec-rate-dis}. 

The Petschek model is the first one that successfully demonstrated rapid energy conversion in the current sheet (CS) system in which the distribution of thermodynamic quantities and the magnetic strength is symmetric with regard to the CS (hereafter, we call it ``symmetric'') without the magnetic shear (hereafter ``shear-less''). It was proposed more than half a century ago and is still considered to be the widely accepted standard model of fast reconnection. However, severe issues in the model have been highlighted. For instance, Nitta (2004) pointed out that the Petschek model is mathematically ill-posed as a Riemann problem. The model remains one degree of freedom. This degree of freedom should justify the concept of the externally driven MRX, i.e., we should be able to arbitrarily set the inflow speed into the diffusion region within a restricted range (e.g., Vasyliunas 1975, Priest \& Forbes 1986). If we consider consistency across all over the reconnection system, we can set a well-posed Riemann problem, and the degree of freedom should be removed (see Nitta 2004). This implies that the Petschek model is incomplete in terms of mathematical problem setting. 

We have noticed another issue in the Petschek model. The most impressive property of the model is the famous ``logarithmic dependence of the reconnection rate on the magnetic Reynolds number.'' This practically implies that the reconnection rate of the Petschek model is approximately a universal constant (of the order of magnitude $\sim 10^{-2}$, known as ``fast reconnection'' in general), almost independent of the magnetic Reynolds number. Currently, the result ``the reconnection rate of the fast MRX is universally $\sim 10^{-2}$'' seems to walk on one's own. However, we should note that this result is based on an unphysical assumption that the thickness of the diffusion region can unlimitedly decrease to adjust for the increase in the magnetic Reynolds number. This is not true because the thickness of the diffusion region exhibits a lower limit, which might be similar to the ion Larmor radius. 

The problem what happens when the above assumption is invalid was solved by Nitta (2007). Herein, we suppose MRX with a sufficiently small magnetic Reynolds number (sufficiently large electric resistivity), in which the diffusion region thickness is larger than the ion Larmor radius. The magnetic diffusion speed is estimated as the magnetic diffusivity divided by the thickness of the diffusion region. The magnetic diffusion speed should balance with the inflow speed in steady models like the Petscheck model. As the magnetic Reynolds number increases, the diffusion region thickness decreases to maintain the balance of the magnetic diffusion speed with the inflow speed. In this scenario, the reconnection rate is almost constant with regard to the variation in the magnetic Reynolds number. This state approximately corresponds to the Petschek model's ``constant universal reconnection rate.'' When the magnetic Reynolds number increases larger than the critical value at which the diffusion region thickness reaches its lower limit, i.e., the ion Larmor radius, the diffusion region thickness cannot decrease any more. Thus, the diffusion speed decreases if the magnetic Reynolds number increases beyond this critical value. The reconnection rate decreases more as the magnetic Reynolds number increases. As a result, the reconnection point at rest is no longer stable and it moves along the CS for higher magnetic Reynolds number cases (Nitta 2007). 

In many cases of the actual MRX phenomena in astrophysical events, the asymmetry of the thermodynamic quantities with regard to the CS and the magnetic shear will be inevitably involved. Therefore, pioneering studies (e.g., the Petschek model) based on the assumption of the shear-less symmetric situation should be generalized to develop a sheared or thermodynamically asymmetric model. Studies on thermodynamically asymmetric MRX, e.g., Petschek \& Thorne (1967), Lin \& Lee (1999), Cassak \& Shay (2007), Birn et al. (2010), and sheared MRX, e.g., Ugai (1981), Soward (1982), and Lin \& Lee (1999) have been conducted. We approach this problem based on an original distinctive point of view: this study focuses on the spontaneous temporal evolution of the MRX that is free from any environmental effect (Section \ref{sec:concept-setup}). This approach intends to simulate a natural spontaneous evolution in a large space. Such evolution should asymptotically reach the self-similar phase (Nitta 2001). We plan to generalize the proposed self-similar MRX model for sheared asymmetric cases. 

Our previous studies on the self-similar MRX model (Nitta et al. 2001, 2002, 2004, Nitta 2006, 2007, Nitta et al. 2016, Nitta \& Kondoh 2019, 2021) demonstrated slower MRX cases than the Petschek model (higher magnetic Reynolds number cases [Figure 4 of Nitta 2007], thermodynamically asymmetric cases about the CS [Figure 8 of Nitta \& Kondoh 2019], sheared magnetic field cases [Figures 10 and 11 of Nitta \& Kondoh 2021]). In these cases, the reconnection rate significantly decreases from a maximum value (prediction of the Petschek model). However, numerical simulations based on the finite difference method face unavoidable inexpedience, i.e., when the diffusion region thickness decreases to a critical value comparable to the mesh size, uncontrollable numerical diffusion dominates in the MRX evolutionary process. In general, such ``numerical MRXs'' proceed faster than the physical MRX (Section \ref{sec:Petschek}). There remains a reasonable doubt whether such numerical MRX can simulate the physically correct dynamical evolution process. 

To resolve the aforementioned issues, we should consolidate an acceptable MRX model that is applicable to general cases, including the slower MRX process. We present the generalized Newtonian (non-relativistic) MRX model in this study and discuss the properties of the spatial structure of the MRX system and the parameter dependence of the reconnection rate. These were independently discussed in our recent studies for the thermodynamically asymmetric (Nitta et al. 2016, Nitta \& Kondoh 2019) or sheared cases (Nitta \& Kondoh 2021). We determined new features of the sheared or thermodynamically asymmetric MRX compared to the shear-less symmetric MRX. For example, the two-layered structure of the jet, the contact discontinuity formation in the plasmoid, the fast shock formation in front of the shorter plasmoid, and the decrease in the reconnection rate for the thermodynamically asymmetric or sheared situation. This study clarifies how the results are altered for the sheared asymmetric cases.

\section{Basic Concept and Numerical Setup} 
\label{sec:concept-setup}
We investigate the asymmetric sheared MRX properties in the self-similar phase (the final phase of the spontaneous temporal evolution). This study is a natural extension of the self-similar MRX model. The model characteristics are summarized in Section 2 of Nitta \& Kondoh (2021) and are omitted from this paper. 

The most important point of the self-similar model is that {\it `this model stands upon keeping consistency across the entire region of the reconnection system. In order to understand the spontaneous determination process of the reconnection system, we must consider not only the region in the vicinity of the X-point (reconnection point) and the reconnection jet, but also the entire region inside the fast-mode (rarefaction) wave front (FRWF) emitted from the X-point at the onset of the reconnection. All changes from the initial equilibrium (e.g., the jet, the plasmoid and the inflow [lobe] region) must be included within the FRWF. Therefore, for example, the inflow speed into the diffusion region is determined by the difference of the total pressure between the diffusion region and the FRWF. If the FRWF exceeds the boundary of the simulation region, the inflow speed in the following evolution cannot be reproduced exactly. Such inflow speed should differ from the correct value obtained from the evolution in the infinitely wide space, and the evolution is affected by artificial boundary condition. This is no longer the ``spontaneous'' evolution in the strict meaning. The prior works paid scant attention to this point. By considering the consistency across the entire region, the reconnection system can determine itself spontaneously'} (Nitta et al. 2016).

\subsection{Simulation code} \label{code}
The numerical setup of the problem we adopted in this study is essentially the same as that in our previous papers (Nitta et al. 2016, Nitta \& Kondoh 2019, 2021); however, slight alterations had been made to simultaneously introduce the asymmetry of the thermodynamic quantities with regard to the CS and the magnetic shear. We consider mainly isothermal current sheet equilibrium with the sheared field, in which the strength of the magnetic field and the distribution of thermodynamic quantities are asymmetric with regard to the CS. For comparison and additional analysis, an isodensity current sheet equilibrium has been considered in a few cases. We have used the OpenMHD code released by Zenitani (``OpenMHD'' \\https://github.com/zenitani/OpenMHD, see Zenitani (2015a)). In the OpenMHD, the magnetohydrodynamics (MHD) equations are solved using the Godunov-type code. Numerical fluxes are calculated using the Harten--Lax-van Leer method for Discontinuities (HLLD) approximate Riemann solver that is coded by Miyoshi \& Kusano (2005). Interpolation of the spatial profiles is given by a minmod limiter. The time-marching is determined by the second-order total variation diminishing (TVD) Runge--Kutta method.

\subsection{Simulation box} \label{sec:sim-box}
The setup of the simulation box is the same as described in Section 2 of Nitta \& Kondoh 2019. To simulate the self-similar phase, which is the terminal phase of the spontaneous temporal evolution, we should prepare a simulation box as large as possible. We demonstrate the essence of the setup again.

We set the Cartesian coordinates $(x, y)$ as shown in Figure \ref{fig:box} for the simulation coordinate. We perform two-dimensional simulations in the $x-y$ plane; hence, the translational symmetry in the $z$-direction is supposed. 

It is necessary to maintain the simulation boundary beyond the fast-mode wave front emitted from the X-point during the simulation to realize spontaneous evolution that is free from any influence of the boundary condition. We must proceed in such a way that the spontaneous temporal evolution approximately reaches the self-similar phase. This implies that we have to prepare simulation box as large as possible. The overview of the simulation box is shown in Figure \ref{fig:box}. The mirror symmetry is imposed on the $y$-axis. The free boundary condition (i.e., the spatial derivative normal to the boundary vanishes) is imposed on the right boundary. The conductive wall boundary condition (the velocity normal to the boundary and the electric field parallel to the boundary vanish) is imposed on the top and bottom boundaries. The conditions set on each boundary are listed in Table \ref{tab:bc}. We must note that we can adopt any boundary condition for the outer boundary; these boundary conditions do not influence the temporal evolution of our simulation. 

\begin{table}[htbp]
\caption{Boundary conditions}
\begin{center}
\begin{tabular}{cl}
\hline
Free boundary & $\partial_x \rho=0$ \\ 
(on $x=2560D$) & $\partial_x P=0$ \\ 
                          & $\partial_x v_x=0$ \\ 
                          & $\partial_x v_y=0$ \\ 
                          & $\partial_x v_z=0$ \\ 
                          & $\partial_x B_x=0$ \\ 
                          & $\partial_x B_y=0$ \\ 
                          & $\partial_x B_z=0$ \\ \hline
Conductive wall boundary & $\partial_y \rho=0$ \\ 
(on $y=\pm 2560D$) & $\partial_y P=0$ \\ 
                          & $\partial_y v_x=0$ \\ 
                          & $v_y=0$ \\ 
                          & $\partial_y v_z=0$ \\ 
                          & $\partial_y B_x=0$ \\ 
                          & $B_y=0$ \\ 
                          & $\partial_y B_z=0$ \\ \hline
Mirror boundary & $\partial_x \rho=0$\\ 
(on $x=0$) & $\partial_x P=0$\\ 
                          & $v_x=0$ \\ 
                          & $\partial_x v_y=0$ \\ 
                          & $v_z=0$ \\ 
                          & $\partial_x B_x=0$ \\ 
                          & $B_y=0$ \\ 
                          & $\partial_x B_z=0$ \\ \hline
\end{tabular}
\label{tab:bc}
\end{center}
\end{table}

The size of the rectangular simulation box is $0 \leq x \leq 2560 D$, $-2560 D \leq y \leq 2560 D$ ($2560 D \times 5120 D$) where $D$ is the half-thickness of the initial CS. This is fairly sufficient to approximately realize the self-similar phase before the fastest wave (i.e., the FRWF) reaches the boundary. This simulation box is resolved by $12800 \times 25600$ meshes (1 mesh=$0.2 D$ in $x$- and $y$-directions) in most cases.

\subsection{Zoom-out Coordinate} \label{sec:zoom-out}
We mainly use the zoom-out coordinate 
\begin{eqnarray}
x' &\equiv x/(V_{A0} t) \label{eq:x'} \\ 
y' &\equiv y/(V_{A0} t)\ .\label{eq:y'}
\end{eqnarray}
to exhibit the global structure in the self-similar phase, where $V_{A0}$ is the Alfv\'{e}n speed in the lower ($y<0$) asymptotic region far distant from the CS and $t$ is the time from the onset of MRX. The self-similar stage is the stationary state if we describe it in the zoom-out coordinate. When we analyze the local structure (e.g., around the X-point), the original simulation coordinate $(x, y)$ is used. 

The zoom-out coordinate is convenient to describe the structure in the self-similar phase. If we plot the self-similarly expanding structure in the zoom-out coordinate, we can easily estimate the expanding or extending speed based on the coordinate values of the structure. For example, a stationary point at the coordinate values ($x', y'$) in the zoom-out coordinate is moving at velocity ($x' \  V_{A0}, y' \  V_{A0}$) in the simulation coordinate.

\subsection{Initial equilibrium} \label{sec:initial-equil}
The magnetic configuration with magnetic shear in the initial equilibrium is schematically shown in Figure \ref{fig:concept}. The degree of the magnetic shear is represented by the shear angle $\theta$ between the magnetic field lines in the upper ($y>0$) and lower ($y<0$) sides of the CS. To introduce the degree of freedom of the azimuthal direction of the X-line relative to the sheared magnetic field lines, the azimuthal angle $\phi$ is defined (Figure \ref{fig:concept}). The azimuthal angle $\phi$ provides the magnetic asymmetry with regard to the CS. 

Moreover, we introduce the asymmetry of the thermodynamic quantities (i.e., the mass density, temperature, and pressure) with regard to the CS. We mainly investigate the isothermal case (i.e., the density and the pressure are asymmetric, maintaining the total pressure balance). As an additional investigation, the isodensity case (i.e., the temperature and pressure are asymmetric) is introduced. The degree of the thermodynamic asymmetry is represented by the ratio $k$ of the magnetic field strength of the lower region ($y<0$) to that of the upper region ($y>0$) (Nitta et al. 2016). 

We set the initial equilibrium for the isothermal case as follows: 
The mass density is 
\begin{eqnarray}
\lefteqn{\rho(y)} \nonumber\\
& = & \rho_0 (1+\cosh^{-2}(y/D)/\beta_0)  \ \ (y<0), \\
& = & \rho_0/(\beta_0 k^2)(\beta_u +\cosh^{-2}(y/D))  \ \ (y>0) \ ,
\end{eqnarray}
the gas pressure is 
\begin{equation}
P(y)=\beta_0 \rho(y)/2 , 
\end{equation}
the velocity is 
\begin{equation}
\mbox{\boldmath$v$}=0 , 
\end{equation}
and the magnetic field is 
\begin{eqnarray}
\lefteqn{\mbox{\boldmath$B$}} \nonumber\\
&=& B_0/k \cdot \tanh(y/D) \cdot [\cos(\phi)  \mbox{\boldmath$\hat{x}$}-\sin(\phi) \mbox{\boldmath$\hat{z}$}] \ \ (y<0), \\
&=& B_0 \tanh(y/D) \cdot [\cos(\phi+\theta)  \mbox{\boldmath$\hat{x}$}-\sin(\phi+\theta) \mbox{\boldmath$\hat{z}$}] \ \ (y>0),
\end{eqnarray}
where $\rho_0$ is the asymptotic mass density, $D$ is the initial (half) thickness of the CS, $\beta_0$ is the asymptotic plasma-beta value in the lower region ($y<0$), $B_0$ is the asymptotic magnetic field strength in the lower region, and \mbox{\boldmath$\hat{x}$} and \mbox{\boldmath$\hat{z}$} are the coordinate bases in the $x$- and $z$-directions, respectively. The word ``asymptotic'' denotes the value in the far distant region from the CS, in which all the quantities are uniform. Our focus is on the low beta cases for the lower region that we frequently encounter in astrophysical applications. Thus, we treat only the case 
\begin{equation}
\beta_0=0.2 \ \ (y<0)
\end{equation}
throughout this study. This is almost the lower limit to perform stable simulation. We select the units of physical quantities as 
\begin{eqnarray}
D=1 \ ,\\
B_0=1 \ ,\\
\rho_0=1 \ .
\end{eqnarray}

\subsection{Resistivity model} \label{sec:res}
We should put resistivity as a model in our MHD simulations. We artificially put a finite resistivity in the rectangular resistive region around the origin (Figure \ref{fig:box}). The resistivity is maintained as a constant independent of the time in each simulation run. It is clarified that in the thermodynamically asymmetric case, the stagnation point (S-point) and X-point will separate and move along the $y$-axis (Nitta et al. 2016). The resistive region should be designed to include this moving X-point during the simulation. We set the resistive region as a vertically elongated shape (Figure \ref{fig:box}) (the shaded region). Based on our experience, we put the size of the resistive region as $D \times 20 D$ ($|x| \leq D$ and $|y| \leq 10D$).

The magnetic Reynolds number is defined as 
\begin{equation}
R_{em} \equiv V_{A0} /(c^2 \eta/D) \ ,\label{eq:Rem}
\end{equation}
where $V_{A0} \equiv B_0/\sqrt{\rho_0}$ is the Alfv\'{e}n speed in the lower asymptotic region, $c$ is the speed of light in vacuum, and $\eta$ is the electrical resistivity (in this study, we adopt the Lorentz--Heaviside units in electromagnetism). We set the magnetic Reynolds number as $R_{em}=24.5$ (this value is plausible to reproduce the Petschek-like MRX; Nitta \& Kondoh 2019) in most cases.

\subsection{Parameter survey}
\label{sec:survey}
We investigate the effect of the magnetic shear and the asymmetry of the thermodynamic quantities on MRX. We fix the the magnetic shear angle as $\theta=60^\circ$, because it is suitable to clearly detect the separable structure of the discontinuities. We adopt the azimuthal angle $\phi$ of the X-line and the thermodynamic asymmetry parameter $k$ as handling parameters. 

We systematically survey the parameter sets as follows: The simulations are executed for the parameter sets of the X-line azimuthal angle $\phi$ from $-80^\circ$ to $+20^\circ$ with an interval step of $10^\circ$ for each thermodynamic asymmetry parameter---$k=1.0, 2.0, 4.0, 8.0, 16.0$ (by fixing the shear angle $\theta=60^\circ$)---for the isothermal initial equilibrium case. 

We focus on the properties of the self-similar phase, which is the final phase of the spontaneous temporal evolution. This requires us to simulate as long as possible. However, some practical circumstances (e.g., total allocated memory, total storage capacity, realistic term period to accomplish this study, and others) limit the appropriate run time. To simulate the temporal evolution completely free from the effect of the simulation boundary, we must stop the simulation run before the fastest signal (i.e., FRWF) emitted from the central region around the diffusion region reaches the boundary. We simulate the temporal evolution after the initial equilibrium is reached up to $t=2400$ for each parameter set. This is the compoundable upper limit to approximately treat the self-similar phase. 

We investigated many sets of parameters $(k, \phi)$; however, several sets exhibited abnormal features. The abnormal termination caused by the unphysical negative pressure is easy to detect because the simulation run is suddenly interrupted. Even if the simulation continues up to $t=2400$, the final state is not always reasonable. The validity of the simulation result is checked as follows: We plot the temporal variation of the reconnection rate (for $t=1700, 1800, 1900, 2000, 2100, 2200, 2300$). Based on our recent numerical studies, we understand that the reconnection rate monotonically decreases in the regular situation of the approximated self-similar phase (see Figure 9 of Nitta \& Kondoh 2021). If the reconnection rate monotonically decreases or decreases as the overall trend, we judge this case as relevant, whereas if the reconnection rate increases as the overall trend, we reject the case.

\section{Results}
\label{sec: results}
We show the results of aforementioned simulations compared to our recent series of studies. The main aspects to be considered are 1) the spatial structure of the outflow, 2) the formation of the forward fast shock (FFS), 3) the plasma mixing (contact discontinuity [CD] formation) in the longer plasmoid, and 4) the reconnection rate. 

There are few new findings in this study. Most of the contributions are modifications or reconfirmation of the achievements in our previous numerical studies on the MRX (Nitta et al. 2016, Nitta \& Kondoh 2019, 2021). We emphasize that this study has been conducted to conclude the series of our studies on the fundamental properties of the thermodynamically asymmetric or sheared MRX. Hence, we set numerical models for the generalized situation, i.e., the asymmetry of the distribution of thermodynamic quantities with regard to the CS and magnetic shear are simultaneously included in the CS system. Thus, this is an extension of our previous studies and is worth to summarize here as the conclusion of the series. 

Let us summarize the achievements of this paper and our previous papers in Table 2. We confirm that some specific properties in the thermodynamically asymmetric case or sheared case hold in the asymmetric sheared case. 

\begin{table}[htp]
\caption{Summary of our numerical studies}
\begin{center}
\begin{tabular}{llll}
\hline
paper & achievement & status & section\\ \hline \hline
N+16 & effects of asymmetry $k \ (\leq 2)$ &&\\ \hline
 & multilayered outflow & new & 3.1\\
 & CD formation & new & 3.2\\
 & FFS formation & new & 3.3\\
 & reconnection rate vs. $k$ & new & 3.6\\ \hline \hline
NK19 & effects of asymmetry $k \ (> 2)$ &&\\ \hline
 & multilayered outflow & reconfirm (N+16) & 3.1\\
 & CD formation & reconfirm (N+16) & 3.1\\
 & FFS formation & reconfirm (N+16) & 3.1\\
 & FFS disappearance & new & 3.3\\ 
 & reconnection rate vs. $k$ & partially new & 3.4\\ \hline \hline
NK21 & effects of magnetic shear $\theta$, $\phi$ &&\\ \hline
 & multilayered outflow & partially new & 4.1, 4.2\\
 & reconnection rate vs. $\theta$, $\phi$ & new & 4.3\\ \hline \hline
NK22 & effects of $k$, $\theta=60^\circ$, $\phi$ &&\\ \hline
 & multilayered outflow & partially new & \ref{sec:outflow} \\ 
 & FFS formation & partially new & \ref{sec:FFS}, \ref{sec:FFS-dis} \\ 
 & CD formation & partially new & \ref{sec:CD}, \ref{sec:CD-formation-application}, \ref{sec:phenomena} \\ 
 & reconnection rate vs. $k$, $\phi$                          & new & \ref{sec:rec-rate}, \ref{sec:rec-rate-dis} \\ \hline
\end{tabular}
\end{center}
\end{table}

Note: The abbreviations denote N+16 (Nitta et al. 2016), NK19, 21 (Nitta \& Kondoh 2019, 2021) in Table 2. We show details of each achievement in the following.

\subsection{Spatial structure of the outflow}
\label{sec:outflow}
First, we show the spatial structure of the outflow (the jet and plasmoid). With regard to our previous studies based on the MHD simulation, we clarify that the Alfv\'{e}n speed $V_{Ax}$ ($\equiv B_x/\sqrt{\rho}$, where $B_x$ is the $x$-component of the magnetic field in the considered side of the CS) projected in the $x$-direction is essential to understand the asymmetrical structure. Figure \ref{fig:ro_d-00023-60-k2} shows the density distribution maps for $k=2$, $\theta=60^\circ$, and $\phi=-80^\circ - +20^\circ$ with an increment $20^\circ$ in the $\phi$ value. The color contour level setting is common to all the panels. We can observe the entire outflow structure and typical discontinuities in the outflow. 

We briefly show the points in Figure \ref{fig:ro_d-00023-60-k2}. The outflow (the jet and two plasmoids) is surrounded by the MHD discontinuities (details are provided below). The extension speed of the plasmoid in the larger $V_{Ax}$ side is more than that in the other side (the lower plasmoid is longer for $\phi=-80^\circ - 0^\circ$, and the upper plasmoid is longer for $\phi=+20^\circ$). We can clearly observe the double claw structure in the longer plasmoid. Note that $V_{Ax}$ is larger in the lower region for $\phi=-80^\circ - 0^\circ$. Thus, the lower plasmoid projects forward from the upper plasmoid. This implies that the lower plasmoid extends faster than the upper plasmoid. The dense upper plasma (yellow small claw) infiltrates the tenuous lower plasma (blue large claw)  for $\phi=-80^\circ - 0^\circ$. The lower plasmoid comprises the yellow small claw embedded in the blue large claw, which exhibits the double claw structure. Inversely, $V_{Ax}$ is larger in the upper region for $\phi=+20^\circ$. The upper plasmoid projects forward from the lower plasmoid. The tenuous lower plasma (blue small claw) infiltrates the dense upper plasma (orange large claw). 

The discontinuities in the jet region are shown in Figures \ref{fig:yprof-60-k2-30}a and \ref{fig:yprof-60-k2+20}a. Figure \ref{fig:yprof-60-k2-30}a shows the profile of the physical quantities at $x'=0.05$ for $k=2$, $\theta=60^\circ$, and $\phi=-30^\circ$, and Figure \ref{fig:yprof-60-k2+20}a shows the profile of the physical quantities at $x'=0.03$ for $k=2$, $\theta=60^\circ$, and $\phi=+20^\circ$. The horizontal axis denotes $y'$ of the zoom-out coordinate in each figure. The structure of the discontinuities in Figure \ref{fig:yprof-60-k2-30}a for $\phi=-30^\circ$ is similar to the shear-less asymmetric case in Nitta \& Kondoh (2019) (see Figure 3 of Nitta \& Kondoh 2019), however, we can observe a new feature arising from the thermodynamic asymmetry. The plasmoid in the lower beta side ($y'<0$) is longer in this case. The jet is surrounded by the SS and intermediate shock (IS), which is followed by the slow-mode expansion fan (SE). We should note that the SE is a property of the thermodynamically asymmetric case. To connect the upper region filled with dense plasma to the lower region filled with tenuous plasma, the SE forms. These structures drive the jet (see $v_x$ profile). The SE further accelerates the jet as the second accelerator following to the IS as the first accelerator. The jet comprises the two-layered structure bounded by the CD between the low-density lower plasma and high-density upper plasma. As shown in Figure \ref{fig:yprof-60-k2+20}a for $\phi=+20^\circ$, the plasmoid in the higher beta side ($y'>0$) is longer. The SS in Figure \ref{fig:yprof-60-k2-30}a alters to the combination of the rotational discontinuity (RD) followed by the SS. This is the result of the magnetic shear to satisfy the coplanarity condition. Note that the $x$-component $B_x$ of the magnetic field is significantly small in the lower side ($y<0$) of the CS in the case of $\phi=+20^\circ$ for geometrical reasons. Thus, the magnetic field line in the lower region outside the jet region should considerably rotate to connect the magnetic field line in the jet region. The RD enables this rotation. This is a notable property of the sheared MRX. In the opposite side, a combination of the IS followed by SE forms (Figure \ref{fig:yprof-60-k2+20}a). To connect the upper region filled with dense plasma to the lower region filled with tenuous plasma, the SE forms. The SE further accelerates the jet as the second accelerator following to the IS as the first accelerator. This is a notable property of the thermodynamically asymmetric MRX. Hence, the hybrid structure that the RD+SS (property of the sheared case) form in one side and IS+SE (property of the thermodynamically asymmetric case) form in another side is an important properly observed in the case with both the thermodynamic asymmetry and the magnetic shear. 

Figures \ref{fig:yprof-60-k2-30}b and \ref{fig:yprof-60-k2+20}b clearly show that the longer plasmoid (longer in the $x$-direction) comprises two plasma components. As shown in Figure \ref{fig:yprof-60-k2-30}b, the lower plasmoid is longer in the $x$-direction. The plasmas coming from the upper region of the CS infiltrate the lower plasmoid to form a CD. As shown in Figure \ref{fig:yprof-60-k2+20}b, the upper plasmoid is longer in the $x$-direction. The plasmas coming from the lower region of the CS infiltrate the upper plasmoid to form a CD. This structure is considered to be a notable property of the asymmetric plasmoid cases as well. We discuss this point in detail in Sections \ref{sec:CD} and \ref{sec:CD-formation-application}.

\subsection{FFS formation}
\label{sec:FFS}
The formation of the long and strong FFS in the thermodynamically asymmetric MRX system was first reported by Nitta et al. (2016). The well-known fast shock in the (shear-less symmetric) MRX system is the reverse fast shock between the reconnection jet and the plasmoid. The scale of the reverse fast shock is similar to the cross section of the jet, which is much smaller than the scale of the FRWF. This reverse fast shock is rather weak. The fast-mode Mach number of the reverse fast shock slightly exceeds unity (in our experience, at most $\sim 1.1$ for the symmetric case; refer to Section 4.3 of Nitta et al. 2016) because the jet speed is limited by the ambient Alfv\'{e}n speed (note we assume the shear-less symmetric case). However, the FFS pointed here is considerably different. The length of the FFS can be similar to the FRWF scale in the side with smaller Alfv\'{e}n speed (the upper side in Figure \ref{fig:pm_d-00023-60-k2-30}a). The fast-mode Mach number of the FFS can significantly exceed unity (with respect to the side with smaller Alfv\'{e}n speed) because the extension speed of the longer plasmoid is larger than the jet speed or the Alfv\'{e}n speed in the side with smaller Alfv\'{e}n speed (the estimation of the fast-mode Mach number is described below). Hence, the FFS is longer and stronger compared with the reverse fast shock we ever know in the MRX system. 

The essence of the FFS formation is the asymmetry of the MHD wave propagation speed across the CS. In the moderately asymmetric system, the fast-mode wave propagation speed in the upper (higher beta) region ($y>0$) is smaller in our problem setting. Assume the extension speed of the lower (lower beta) plasmoid exceeds the fast-mode propagation speed in the upper (higher beta) region. When the fast-mode compression wave produced around the lower (lower beta) plasmoid tip propagates across the CS, it immediately forms the shock wave. For example, in the isothermal asymmetric case ($k=2$ without magnetic shear), the fast-mode Mach number of the FFS is $\sim 1.6$ (as the largest value near the lower plasmoid tip; refer to Figure 11 of Nitta et al. 2016). In the highly asymmetric case, the lower plasmoid extension speed reduces to be sub-fast for the upper (higher beta) plasma, and the FFS is not formed (Figures 5 and 7 of Nitta \& Kondoh 2019).  

In the isothermal sheared case (without the asymmetry of the thermodynamic quantities, i.e., $k=1$), we also find the projection or preceding of the plasmoid in one side if $V_{Ax}$ is asymmetric with regard to the CS. However, the FFS is not formed because the fast-mode propagation speed in the $x$-direction is almost symmetric, and the extension speed of the projected plasmoid is sub-fast with regard to the opposite side plasma (Section 5.2.4 of Nitta \& Kondoh 2021). 

Does the FFS form in the thermodynamically asymmetric sheared case? The answer is yes, because the essential factors of the FFS formation are asymmetry of the fast-mode propagation speed and projection (preceding) of one-side plasmoid from another plasmoid (the extension of the one-side plasmoid is faster than another in the thermodynamically asymmetric case). Figure \ref{fig:pm_d-00023-60-k2-30}a shows the magnetic pressure distribution for $k=2, \theta=60^\circ, \phi=-30^\circ, t=2300$. The blue--black thin structure along the CS ($y=0$, around $x' \sim 0.1-0.2$) is the plasmoid. The hump structure (the blue--black round structure) around $x' \sim 0.15$ is the upper plasmoid. The thin and elongated structure between $x' \sim 0.15-0.53$ is the lower plasmoid. Thus, the lower plasmoid extension speed is $\sim 0.53 V_{A0}$, whereas the fast-mode speed in the upper region is $\sim 0.41  V_{A0}$ in this case. The fast-mode Mach number in the upper region of the lower plasmoid extension speed is $\sim 1.3$. This value directly denotes the largest value of the fast Mach number of the FFS because the plasma in front of the FFS is almost at rest. The shock strength is rather weaker than the shear-less case (the fast Mach number $\sim 1.6$ for $k=2$, see Figures 11 and 12 of Nitta et al. 2016). The fast Mach number of the FFS monotonically decreases along the FFS from the contact point with the CS to upward direction. We can observe a clear FFS in front of the upper plasmoid; however, its strength is very small (Figure \ref{fig:pm_d-00023-60-k2-30}b: profile at $y'=0.005$). In the sheared case, the magnetic energy caused from the sheared component remains in the reconnection outflow. The reconnection outflow slows compared with the shear-less case. Hence, the strength of the FFS decreases.

\subsection{CD formation in the plasmoid}
\label{sec:CD}
An interesting and important property of the thermodynamically asymmetric case is the formation of the CD in the longer plasmoid (Figure 8 of Nitta et al. 2016). The plasmas in the upper and lower side of the CS are separated by the CS in the initial state. These components never mix in the MHD framework because they exist on different magnetic field lines. However, once the MRX evolves, the plasmas from both sides of the CS coexist on the same magnetic field lines in the same plasmoid. This structure is ideal for effective plasma mixing when the plasmoid structure crushes. This CD formation was first reported by Nitta et al. (2016). In the case of the asymmetry of the thermodynamic quantities, the plasmoid structure is asymmetric with regard to the CS. The plasmoid in the lower beta side projects forward (in the $x$-direction) from the higher beta side one. This implies that the lower beta side plasmoid extends faster than the other plasmoid. The inflow from the higher beta side dominates to maintain the magnetic influx balance. Based on the momentum balance, the reconnection jet inclines toward the higher beta side. The inclined jet obliquely hits the CS. Thus, the non-negligible fraction of the higher beta side plasma is caught up into the lower beta side plasmoid to form the new CD. 

Moreover, we expect the CD formation in the asymmetric sheared MRX. In fact, we can find the CD in Figure \ref{fig:ro_d-00023-60-k2} denoting the mass density distribution. The maps of Figure \ref{fig:ro_d-00023-60-k2} for $\phi=-80^\circ - 0^\circ$ show the case $k=2, \theta=60^\circ, t=2300$. The entire plasmoid exhibits an asymmetric crab-hand shape. The lower claw projects forward from the upper one in the $x$-direction (the lower plasmoid extends faster than the upper plasmoid). The lower claw exhibits the double structure: The plasma coming from the upper (lower beta) region (in $\phi=-20^\circ$, the orange region around $0.16<x'<0.34$) infiltrates the lower plasmoid that comprises the lower (higher beta) plasma (the blue region around $0.16<x'<0.49$). The boundary between the orange and blue region defines the CD, which was first pointed by Nitta et al. 2016. 

Though it may seem paradoxical, the case that the lower beta side plasma infiltrates the higher beta side plasma is feasible in the sheared case, even if the Alfv\'{e}n speed is larger in the lower beta side. Figure \ref{fig:ro_d-00023-60-k2} for $\phi=+20^\circ$ shows the aforementioned case ($k=2, \theta=60^\circ, \phi=+20^\circ, t=2300$). Although the Alfv\'{e}n speed is larger in the lower region, the upper plasmoid projects forward from the lower plasmoid (the upper plasmoid extends faster than the lower plasmoid) in this case because $B_x$ in the lower region is very small. As the Alfv\'{e}n speed in the $x$-direction ($V_{Ax}$) is smaller in the upper (higher beta) region, the entire length of the plasmoid is very short. We can observe that the plasma coming from the lower region (blue) infiltrates the upper plasmoid to form the inner claw ($0.1<x'<0.12$); however, the total amount of the plasma ingress is very small compared to the case of Figure \ref{fig:ro_d-00023-60-k2} for $\phi=-40^\circ - 0^\circ$. In front of this inner claw, the plasma coming from the upper region (yellow and orange) exists ($0.1<x'<0.17$). This double claw structure is essentially the same as that in the case of Figure \ref{fig:ro_d-00023-60-k2} for $\phi=-80^\circ - 0^\circ$. 

Note that the magnetic field lines in the longer plasmoid thread the CD (Figures \ref{fig:ro_d-00023-60-k2up}a and \ref{fig:ro_d-00023-60-k2up}b). This implies that the plasmas coming from both sides of the CS lie on the same magnetic field lines (Nitta et al. 2016) in the same plasmoid. Such a situation enables effective plasma ingress and plasma mixing as in the shear-less asymmetric case.

\subsection{Reconnection rate}
\label{sec:rec-rate}
MRX is important as the energy converter. The conversion power can be estimated by the reconnection rate and the spatial scale of the system. We demonstrate the parameter dependence of the reconnection rate on the thermodynamic asymmetry $k$ and X-line azimuthal angle $\phi$ fixing the shear angle as $\theta=60^\circ$ here. 

The reconnection rate is defined as the normalized reconnection electric field. We must note here that, in the thermodynamically asymmetric or sheared case, the reconnection point (the X-point) is moving along the $y$-axis, in general. Thus, the reconnection electric field should be measured in the X-point comoving frame (see appendix of Nitta \& Kondoh 2021). The reconnection rate is defined as follows: 
\begin{equation}
R^* \equiv |E'_z|/(V_{A0} B_0) = \eta |j_z^*|/(V_{A0} B_0), \label{eq:R*}
\end{equation}
where $E'_z$ is the $z$-component of the electric field estimated in the X-point comoving frame, $j_z^*$ is the $z$-component of the electric current density at the X-point, $V_{A0}$ and $B_0$ are the Alfv\'{e}n speed and magnetic field strength in the lower asymptotic region far distant from the CS (the magnetic field strength and density in the lower region do not alter during the parameter survey). Note that this reconnection rate is a Galilean invariant. 

In the ideal situation, the reconnection rate $R^*$ gradually decreases as time proceeds and asymptotically settles to a terminal value. As we mentioned in Section \ref{sec:survey}, the time variation of $R^*$ is not monotonical in several cases of our actual simulations. We estimate the approximated terminal value by the temporal average of the values at $t=$1700, 1800, 1900, 2000, 2100, 2100, 2200, and 2300. We can remove unfavorable (probably numerical fake) irregular temporal oscillation using this temporal averaging treatment. 

The parameter dependence of the reconnection rate is shown in Figure \ref{fig:logR*60k1-2-4-8-16-ave}. We performed a systematic survey of the parameters denoting the thermodynamic asymmetry and magnetic shear. To the best of our knowledge, this is the first such survey in the history of MRX research. The shear parameter is fixed to a typical value of $\theta=60^\circ$ because the influence of $\theta$ is moderate and easy to expect (Figures 10 and 11 of Nitta \& Kondoh 2021). The dependence with respect to the thermodynamic asymmetry parameter $k$ ($k=1, 2, 4, 8, 16$) and the X-line azimuthal angle $\phi$ ($-80^\circ \leq \phi \leq +20^\circ$ at intervals of $10^\circ$) are shown. Data points are denoted by dots. Note that when $\phi=-90^\circ, +30^\circ$, $R^*=0$ the annihilable $x$-component of the magnetic field vanishes on either side. Few data points are omitted owing to the abnormal termination of the simulation run or unphysical strange behavior of the temporal evolution (Section \ref{sec:survey}). 

The reconnection rate $R^*$ decreases as the thermodynamic asymmetry $k$ increases as the overall trend. This feature was already pointed by Nitta \& Kondoh (2019). As we expected from the magnetic configuration ($B_x \rightarrow 0$ as $\phi \rightarrow -90^\circ, +30^\circ$), $R^*$ significantly decreases in both ends of this figure. However, we find a new feature that the decrement in the left side ($\phi \sim -90^\circ$) is significant compared with that on the right side ($\phi \sim +30^\circ$). Discussion for the physical process of this left--right asymmetry is made in Section \ref{sec:rec-rate-dis}. We should realize that variation in the thermodynamic asymmetry parameter $k$ and the azimuthal angle $\phi$ of the X-line may alter the reconnection rate $R^*$ in several orders of magnitude.

\section{Summary and Discussion}
\subsection{Summary} 
\label{sec:sum}
As an inevitable result of the spontaneous evolution that is free from any environmental influence, the MRX system asymptotically tends to the self-similar phase (Nitta et al. 2001). Our aim is to clarify the nature of spontaneous MRX in the self-similar phase as an elementary process of single MRX system. This is the self-similar model describing the spontaneous spatial evolution from the scale of the order of the ion Larmor radius to the flare loop scale through self-similar expansion. 

This study aims at concluding the series of our numerical studies on thermodynamically asymmetric or sheared MRX as an elementary process of plasma astrophysical phenomena (Nitta et al. 2016, Nitta \& Kondoh 2019, 2021). To comprehensively summarize the series, we considered a generalized situation for investigation. We studied the effects of the asymmetry of the thermodynamic quantities with regard to the CS and magnetic shear on the spontaneously evolving MRX system. We focused on the terminal state of the spontaneous temporal evolution, i.e., the self-similar phase. This study is a natural extension of our previous studies on the ``self-similar MRX model'' (Nitta et al. 2001, 2002, Nitta 2004, 2006, 2007, Nitta et al. 2016, Nitta \& Kondoh 2019, 2021) with regard to the fundamental properties of the spontaneous MRX. To clarify the structural and other properties of the self-similar MRX system, we performed two-dimensional MHD numerical simulations of the quasi-terminal phase of the long-term temporal evolution from the initial current sheet equilibrium. We adopt an approximate Riemann solver (HLLD code) to maintain sufficient resolution for the complex discontinuity structure. 

Note that, in this study, the shear component of the magnetic field is induced by the azimuthal rotation of the initial magnetic configuration about the $y$-axis (perpendicular to the initial CS). This relative rotation is represented by the shear angle $\theta$. Hence, the magnetic field strength holds in the parameter survey of $\phi$ and $\theta$. In most cases, the magnetic shear is fixed to a typical value of $\theta=60^\circ$ because it is plausible to clearly detect the separable structure of discontinuities, e.g., RD and SS. With regard to the initial equilibrium state, the magnetic field strength and thermodynamic quantities in the lower half region ($y<0$) hold in all the parameter sets that we survey. It is important to keep normalization of the physical quantities among all the parameter sets. 

We contributed minimal new findings in this study (Table 2) because the main aim was to unify the effects of the thermodynamic asymmetry and magnetic shear. Most properties pointed in this study relate to the modification or reconfirmation of the results of the thermodynamically asymmetric (Nitta et al. 2016, Nitta \& Kondoh 2019) or sheared MRX systems (Nitta \& Kondoh 2021). We clarify how and why known properties in our previous studies change under thermodynamically asymmetric and the magnetically sheared conditions. Reconfirmation is worth as well because it clarifies that the properties of the thermodynamically asymmetric MRX or the sheared MRX also appear in the asymmetric sheared MRX. To prove this, we conducted a systematic parameter survey on the thermodynamic asymmetry and magnetic shear for quantitative analyses. 

We present the results of this study in comparison to our previous studies: 1) formation of the FFS in front of the shorter plasmoid in the higher beta region, 2) formation of the CD in the longer plasmoid, 3) parameter dependence of the reconnection rate. We discuss the physical and phenomenological properties of each result in the following section.

\subsection{FFS formation}
\label{sec:FFS-dis}
The long and strong FFS in front of the shorter plasmoid was first pointed by Nitta et al. (2016) for the shear-less asymmetric cases. We reconfirmed the formation of the FFS for the sheared asymmetric cases in this study. The FFS is fascinating as a new particle acceleration site. The FFS forms under the condition that the extension speed of the longer plasmoid is larger than the fast-mode propagation speed in the opposite side of the CS (Figures 5 and 7; Section 3.3 of Nitta \& Kondoh 2019). For the shear-less asymmetric case (Nitta et al. 2016, Nitta \& Kondoh 2019), the FFS can be rather strong (fast-mode Mach number $\sim 1.6$ for the maximum value). On the other hand, for the sheared asymmetric case (in this study), the long FFS is also formed if the formation condition is satisfied as in the shear-less asymmetric case; however, it is rather weaker than the shear-less case (Figure \ref{fig:pm_d-00023-60-k2-30}b; the fast Mach number $\sim 1.3$ in this case). 

The reason is in the magnetic shear. Remember that the annihilable component of the magnetic field is only the $x$-component (Figure \ref{fig:concept}). In the sheared case, the magnetic field inevitably includes the $z$-component that cannot be annihilated. This implies that only a part coming from the $x$-component of the initial magnetic energy is reducible by the MRX. Therefore, the outflow speed is smaller than that in the shear-less case. Moreover, the extension speed of the plasmoid is smaller. Thus, the resultant FFS is weaker than that in the shear-less case. It is trivial that if the shear angle $\theta$ increases while maintaining the normalization mentioned in Section \ref{sec:sum}, the reconnection outflow speed decreases; thus, the strength of the FFS decreases. The FFS will disappear at last at a critical value of $\theta$ depending on $k$ and $\phi$ (we do not obtain the concrete functional dependence owing to the available limit of computational resources).

\subsection{CD formation in the plasmoid}
\label{sec:CD-formation-application}
The thermodynamic asymmetry parameter $k$ or the azimuthal direction $\phi$ of the X-line affects the extension speed of the plasmoids in each side of the CS. Thus, there are two causes of the asymmetry with regard to the CS; the asymmetry caused by the asymmetric distribution of the thermodynamic quantities (thermodynamic asymmetry) that is characterized by $k$ and the asymmetry caused by the asymmetric magnetic shear structure (magnetic asymmetry) that is characterized by $\phi$. We do not distinguish these two kinds of asymmetry in this section. We discuss the influence from these two types of asymmetry in the following. If the plasmoid extension speed is asymmetric, plasmas coming from the side of the shorter plasmoid infiltrate the longer plasmoid (Nitta et al. 2016, Nitta \& Kondoh 2019). In this case, the longer plasmoid exhibits a double crab-hand claw structure (Figures \ref{fig:ro_d-00023-60-k2up}a and  \ref{fig:ro_d-00023-60-k2up}b). The small claw comprises the plasma component coming from the opposite side of the CS. This small claw is embedded in the large claw, which includes the plasma component of this side. These two plasma components in the double claw structure are bounded by the CD.  

The formation process of the CD was clarified in Nitta et al. (2016) (refer to Section 4.2 and Figure 9 of Nitta et al. 2016) for the thermodynamic asymmetry. The characteristic structure is an important property of the thermodynamically asymmetric MRX systems. The reconnection jet from the X-point mainly comprises the plasmas from the side with smaller $B_x$ because of the magnetic influx balance (note that the annihilable component is only the $B_x$). The CS between the two plasmoids is lifted and tilted by the fast-mode compression wave, which is emitted from the longer plasmoid tip. The plasma from the smaller $B_x$-side in the reconnection jet strikes the CS from an offset direction to push the tilted CS upward. Then, a non-negligible fraction of the smaller $B_x$-side plasma in the jet infiltrates the plasmoid in the opposite side to form a CD. Consequently, the smaller and larger $B_x$-side plasmas coexist in the longer plasmoid threaded by the same reconnected magnetic field lines.

The CD formation is commonly observed in the shear-less asymmetric ($k>1$) MRX system (Nitta et al. 2016, Nitta \& Kondoh 2019). In such a case, the plasmoid in the side with larger Alfv\'{e}n speed exhibits a double claw structure. We also find a similar structure in the sheared symmetric ($k=1$) MRX system (Nitta \& Kondoh 2021) for the magnetic asymmetry. In this case, the double claw structure forms in the plasmoid with larger $B_x$ side. However, in the sheared case, a different situation can occur. The deterministic factor is the projected Alfv\'{e}n speed $V_{Ax}$ in the $x$-direction. See the map for $\phi=+20^\circ$ in Figure \ref{fig:ro_d-00023-60-k2}. If $V_{Ax}$ is larger in the upper region, the double claw structure is formed in the upper plasmoid, even though the Alfv\'{e}n speed is larger in the lower region. This is a new finding of this study. 

Considering the phenomenological importance, the CD formation pointed here is directly related to the ingress of plasma across the CS from one side to another (Section \ref{sec:phenomena}). The plasmas in both sides of the CS are initially separated by the CS. Hence, these plasmas can never be mixed in the ideal MHD framework. The plasma in the side with smaller $V_{Ax}$ infiltrates the side with larger $V_{Ax}$ via the asymmetric sheared MRX. The area of the infiltrated plasma (area of the smaller claw of the double claw structure), as shown in Figure \ref{fig:ro_d-00023-60-k2}, exhibits the intensity of plasma ingress. As the structure expands self-similarly with time, the area indicates the amount of the plasma ingress per unit length in $z$-direction along the X-line. Although performing quantitative analysis is difficult because we cannot objectively determine the area of the smaller claw, we understand the qualitative tendency of the plasma ingress, as shown in Figure \ref{fig:ro_d-00023-60-k2}. Considering together with the characteristics of the reconnection rate, as shown in Figure \ref{fig:logR*60k1-2-4-8-16-ave}, the maximum of the area of the plasma ingress approximately coincides with the maximum of the reconnection rate (e.g., $\phi \sim -20^\circ- -30^\circ$ for $k=2, \ \theta=60^\circ$). 

We point an important fact that, once asymmetric MRX occurs, the plasmas from one side (with smaller $V_{Ax}$) of the CS flow over the CS, and infiltrate the other side (with larger $V_{Ax}$) of the CS. The total amount of the plasma ingress is in proportion to the square of the time from the onset of MRX because the system is self-similarly expanding. We must note that these two components of plasmas coexist on the same reconnected magnetic field lines in the longer plasmoid. Such structure seems to be ideal for realizing efficient plasma mixing. 

On the other hand, in the symmetric MRX, plasmas initially in one side of the CS stay in the same side even MRX proceeds. As the result, the lower/upper plasmoid consists of the plasmas initially in the lower/upper side of the CS, respectively. These two plasma components are spatially separated and cannot be mixed in the MHD frame work. With regard to the kinetic (particle-scale) perspective, the particles of the two plasma components divided in the upper and lower plasmoid can be mixed only by the diffusion process based on the individual particle motion along reconnected magnetic field lines. The spatial distance between two plasmoids increases in proportion to the time as the self-similar expansion proceeds. Resultantly, the diffusion speed (plasma flux per unit area) decreases in inversely proportion to the time. The area across that the diffusion carries the plasma increases in proportion to the time as self-similar expansion proceeds. Resultantly, the amount of the plasma ingress per unit time that can be carried by the diffusion is independent of the time ($\propto t^0)$. 

We stress that the plasma mixing in the asymmetric MRX is more rapid than that in the symmetric MRX. Both the upper and lower plasma components coexist in the longer plasmoid in the asymmetric case. This is not the result of diffusion, but of fluid ingress across the CS. These two plasma components coexist on the same reconnected magnetic field lines in the longer plasmoid. The plasma mixing across the CS can be carried also by the plasma fluid infiltration in addition to the particle diffusion. The amount of the plasma ingress by the fluid infiltration increases in proportion to square of the time in the self-similar phase that is expanding at the fast-mode propagation speed. Thus, the plasma ingress per unit time by the fluid infiltration increases in proportion to the time ($\propto t^1$). 

The CD formation in the longer plasmoid and the plasma mixing associated with the plasma ingress is a new finding of our series of studies on asymmetric self-similar MRX. In the asymmetric MRX, the plasma mixing across the CS is carried not only by the particle diffusion, but also by the fluid infiltration. By comparing these two processes, the amount of the plasma ingress as the fluid infiltration in the asymmetric MRX becomes dominant as time proceeds. We can understand that the asymmetric MRX is effective for plasma mixing.

\subsection{Reconnection rate}
\label{sec:rec-rate-dis}
The reconnection rate is an important index that denotes the speed of the magnetic energy conversion. We are interested in the variation in the reconnection rate based on the orthodox symmetric shear-less case. The aim of our studies is unique and independent of the recent trend of studies. We compare our studies with other recent studies. 

We describe the recent trend of MRX studies so-called turbulent MRX and plasmoid mediated/induced MRX. These studies employ (approximately) uniform small resistivity in the CS system. Such system is unstable, e.g., for the tearing mode. Many small-scale non-steady MRXs are driven by the background turbulent velocity field in the turbulent MRX model (Lazarian \& Vishniac 1999). Many small tearing islands spontaneously occur in the plasmoid mediated MRX model (Loureiro \& Uzdensky 2016). The interacting many MRXs enhance each other. To achieve an enormous magnetic energy release, a large number of small MRXs should work in these models. The resultant reconnection rate is almost independent of the magnetic Reynolds number (this situation is called ``fast MRX'' after a fashion of certain researchers). As related topics, we refer to the sporadic Petschek-type MRX, plasmoid induced MRX, and fractal MRX for reference. Shibayama et al. (2015, 2019) pointed that sporadic small-scale Petschek-type MRXs occur in the CS system even with uniformly distributed small electric resistivity. Shibata (1996) and Shibata \& Tanuma (2001) reported that the ejection of the plasmoid by the first MRX thins the remaining CS and triggers the following faster secondary MRX (plasmoid-induced MRX). This CS-thinning process with the tearing instability recurrently continues until the CS thickness reaches the scale similar to the ion Larmor radius. The CS system with multiple MRXs decays to form a fractal structure (fractal MRX). The recent trend of the fast MRX seems to move to the complicated system with multiple MRXs. 

We are interested in another type of fast reconnection originated by the Petschek model (Petschek 1964) that is characterized by localized resistivity and impressive long figure-X shaped slow shocks (SS) as the main magnetic energy converter. The magnetic energy conversion by a single MRX is important in this type of MRX. The spatial scale of this MRX system can grow unlimitedly, exceeding the initial CS thickness. Thus, the single MRX can release an enormous amount of magnetic energy in such system. We must note that the total power of the magnetic energy conversion is estimated not only by the reconnection rate, but approximately estimated by the product, i.e., (reconnection rate)$\times$(MRX system spatial scale)$\times$(number of MRX systems). Recent vogue studies on turbulent MRX or plasmoid-mediated MRX with multiple MRX systems remarked on the effect of (reconnection rate)$\times$(number of MRX systems). On the other hand, our study on the self-similar MRX notices the effect of (reconnection rate)$\times$(MRX system scale). The single MRX started in the scale of the order of the ion Larmor radius explosively expands self-similarly at fast-mode propagation speed, maintaining the reconnection rate of the Petschek model in the maximum case. The reconnection rate of the self-similar model depends on the magnetic Reynolds number (see Nitta 2007); however, its value is $\sim 10^{-2}$ at the maximum. Thus, we can define it as ``fast'' MRX. 

The reduction in the reconnection rate caused by the asymmetry of the distribution of thermodynamic quantities with regard to the CS was first systematically investigated in our previous studies (Nitta et al. 2016 and Nitta \& Kondoh 2019). On the other hand, the reduction in the reconnection rate caused by magnetic shear has been reported in many previous literatures. For example, studies based on three-dimensional MHD simulation for Earth's day-side MRX phenomena reveal the parameter dependence of the reconnection rate (e.g., Lee et al. 2002, Borovsky et al. 2008, Komar \& Cassak 2016). Especially, the approach in Komar \& Cassak (2016) (the effect of the rotation of the solar wind (the magnetosheath) side magnetic field direction) is very similar to our study. Similar to our result, the decrease in the reconnection rate by the magnetic shear was reported. However, their interest focused on the actual Earth's day-side phenomena; thus, the range of the variation in controlling parameters is restricted for the approximated geomagnetospheric case. Aside from these studies motivated by phenomenological interest, our studies focus on clarifying the fundamental properties of the sheared asymmetric MRX as an elementary process of plasma astrophysics. As per our knowledge, for the first time, Nitta \& Kondoh 2021 attempted to reveal the systematic parameter dependence of the reconnection rate on the magnetic shear that is parametrized by the shear angle $\theta$ and azimuthal angle $\phi$ of the X-line. 

The parameter survey of our study is highly systematic. This study extends the parameter survey for the thermodynamic asymmetry $k$ and azimuthal angle $\phi$ fixing the shear angle $\theta=60^\circ$ as a representative. The systematic survey with three parameters, i.e., $k$, $\theta$, and $\phi$, is unrealistic owing to limited computational resources. 

The shear angle $\theta$ varies in $0 \leq \theta < \pi$ (Nitta \& Kondoh 2021). This is the entire range of $\theta$ where MRX occurs. We have fixed $\theta=60^\circ$ in this study because based on our experience, discontinuities properly formed in the sheared MRX system are well separated in this case. The dependence on $\theta$ is moderate and monotonic (Figure 10 of Nitta \& Kondoh 2021). The effect of the variation in $\theta$ is easy to expect because the annihilable component $B_x$ monotonically decreases as $\theta$ increases. The situation should be the same in the sheared asymmetric cases. 

The azimuthal angle $\phi$ can vary in the range of $-90^\circ$ -- $+30^\circ$. We survey $\phi=-80^\circ$ -- $+20^\circ$ at intervals of $10^\circ$ (note MRX does not occur when $\phi=-90^\circ$ and $+30^\circ$ because the annihilable magnetic component $B_x$ in one side vanishes). Thus, we cover the entire range of $\phi$. 

The thermodynamic asymmetry $k$ can vary in the range $1 \leq k < \infty$. The case $k=1$ was confirmed in Nitta \& Kondoh (2021). It is impossible to survey up to infinity. We survey $k=2, 4, 8, 16$. The MRX drastically slows as $k$ increases (Figure \ref{fig:logR*60k1-2-4-8-16-ave}). The result monotonically changes with $k$. Thus, we can easily expect the result with larger $k$. We believe that our systematic parameter survey would be sufficient to understand the general properties of the sheared asymmetric MRX. 

Although we can consider many types of the initial equilibrium, it is impossible to test all the cases. This study mainly considers the isothermal initial equilibrium. Complementary to this, we have considered a few supplementary isodensity initial equilibrium cases to understand the effect of temperature (i.e., the sound speed) asymmetry. We cannot find any essential difference between them because the essential factor is the difference in the propagation speed of MHD fundamental wave modes across the CS. 

Totally, we think that our parameter survey in our series of studies is approximately complete to understand the fundamental properties of the sheared asymmetric MRX. We clarified the parameter dependence of the reconnection rate in a wide variety of the plasma-beta case (Nitta 2004), the magnetic Reynolds number case (Nitta 2006, 2007), the thermodynamically asymmetric case (Nitta et al .2016, Nitta \& Kondoh 2019), and the sheared case (Nitta \& Kondoh 2021) in our previous studies. These are important results of the ``self-similar MRX model'' (Nitta et al. 2001, 2002, Nitta 2004). The most important contribution of these studies of the self-similar MRX model is the clarification on the significant variation of the reconnection rate even in the Petschek-like fast MRX system (see a brief summary in appendix \ref{sec:app-A}). 

Also in the asymmetric sheared system, the reconnection rate considerably varies depending on the parameters. The thermodynamic asymmetry parameter $k$ influences on the reconnection rate as a power law dependence (Figure \ref{fig:logR*60k1-2-4-8-16-ave}). This has been already clarified in our previous study (Nitta \& Kondoh 2019) on the asymmetric shear-less case. The result of the present study is consistent with the result of that paper. 

However, the dependence on the azimuthal angle $\phi $ of the X-line is rather different from the result of Nitta \& Kondoh (2021) for the symmetric sheared system. Figure \ref{fig:logR*60k1-2-4-8-16-ave} clearly shows the left--right asymmetry of the response of the reconnection rate to $\phi$. This is the case for $\theta=60^\circ$. Thus, the geometrically symmetric direction of the X-line is $\phi=-30^\circ$ (the mid-angle of the sheared magnetic field lines). The reconnection rate considerably decreases as $\phi$ varies from its symmetric value $\phi=-30^\circ$. This variation is symmetric about $\phi=-30^\circ$ only in the case $k=1$ (this had been already demonstrated in Nitta \& Kondoh 2021). The decrease in the left half ($\phi < -30^\circ$) is more significant than that in the right half ($\phi > -30^\circ$). We discuss this tendency below. 

Remember that the annihilable magnetic component in our problem is only the $x$-component $B_x$. Moreover, in the self-similar phase, the distribution of the physical quantities in a finite area around a fixed point in the zoom-out coordinate is stationary (note that the self-similar solution is stationary in the zoom-out coordinate). Hence, in the stationary problem around the diffusion region, the magnetic influx of $B_x$ into the diffusion region should be balanced between the lower and upper regions. First, we discuss the isothermal case using a simple toy model. The $x$-component of the magnetic field in the upper and the lower sides are
\begin{equation}
B_{xu}=\frac{B_0}{k} \cos \phi \ (y>0) \label{eq:bxu}
\end{equation}
and 
\begin{equation}
B_{xd}=B_0 \cos (\theta+\phi) \ (y<0) ,  \label{eq:bxd}
\end{equation}
respectively. The mass density in both sides are
\begin{equation}
\rho_u=\frac{\rho_0 (k^2(1+\beta_0)-1)}{k^2 \beta_0} \ (y>0)
\end{equation}
and
\begin{equation}
\rho_d=\rho_0 \ (y<0) ,
\end{equation}
respectively. The magnetic influx from the upper and lower regions should be balanced, 
\begin{equation}
B_{xu} \ v_{inu}=B_{xd} \ v_{ind} , \label{eq:mag-influx-balance}
\end{equation}
where $v_{inu}$ and $v_{ind}$ denote inflow speed into the diffusion region from the upper region and the lower region, respectively. Thus, the mass influx ratio $r_m$ of the upper one to the lower one is
\begin{equation}
r_m \equiv \frac{\rho_u \ v_{inu}}{\rho_d \ v_{ind}}=\frac{k^2(1+\beta_0)-1}{k \beta_0}\frac{\cos (\theta+\phi)}{\cos \phi} .
\end{equation}
We plot the $\phi$-dependence of the ratio $r_m$ for each thermodynamic asymmetry parameter $k=1, 2, 4, 8, 16$ in Figure \ref{fig:rm-isothermal}a providing $\beta_0=0.2$, $\rho_d=1.$ and $\theta=60^\circ$ as in our simulation. This $r_m$ directly shows the mass ratio of the upper plasma component to the lower one in the reconnection outflow. We can find that the plasma from the upper region is dominant in the outflow for the thermodynamically asymmetric cases $k>1$. Note that these two components of plasmas are not mixed in the outflow and separately form a two-layered structure (Figure 3 of Nitta et al. 2016). In addition, the upper plasma component is also dominant as $\phi$ approaches $-90^\circ$. As the upper component in the outflow becomes more dominant, the outflow speed decreases more because the mass density is larger and the releasable magnetic energy in the inflow is smaller in the upper component. If the outflow speed is suppressed, the inflow speed is also suppressed owing to the suppression of the FRW emission (remember that the inflow is induced by the FRW emission in the spontaneous MRX system). Thus, the reconnection rate is suppressed more for the larger $k$ or when $\phi$ tends to $-90^\circ$. 

The aforementioned analysis is also applicable to the isodensity case. We think that the isodensity case is not realistic in actual astrophysical problems because there might be no relaxation process to attain the isodensity equilibrium. However, consideration of and comparison with the isothermal case may be helpful to effectively understand the reconnection-rate-determining process. 

The annihilable magnetic component is the same as defined by equations (\ref{eq:bxu}) and (\ref{eq:bxd}). The mass density is 
\begin{equation}
\rho_u=\rho_d=\rho_0 .
\end{equation}
Based on the magnetic influx balance (\ref{eq:mag-influx-balance}), we obtain the mass influx ratio $r_m$
\begin{equation}
r_m=k \frac{\cos (\theta+\phi)}{\cos \phi} .
\end{equation}
We plot the $\phi$-dependence of the mass flux ratio $r_m$ for the isodensity case in Figure \ref{fig:rm-isothermal}b (for the same other parameters shown in Figure \ref{fig:rm-isothermal}a). We can derive a similar trend as that of the isothermal case. As the upper component in the outflow becomes more dominant, the outflow speed decreases more because the releasable magnetic energy in the inflow becomes smaller in the upper component. Note that $r_m \propto k^1$ for the isodensity case, whereas approximately $r_m \propto k^2$ for the isothermal case. Thus, the decrease in the reconnection rate for large $k$ in the isodensity case should be moderate compared with the isothermal case. 

We can observe this tendency in Figure \ref{fig:R*-phi-60-k16-isod_isot}. We show the $\phi$-dependence of the reconnection rate (the isothermal case (solid curve) for $\theta=60^\circ$, $k=16$ and the isodensity case (dashed curve) for $\theta=60^\circ$, $k=16$). We can observe the same tendency in both cases; however, the reconnection rate is smaller in whole for the isothermal case as discussed earlier. 

Figure \ref{fig:logR*60k1-2-4-8-16-ave} clearly shows that the reconnection rate sensitively depends on the azimuthal direction $\phi$ of the X-line; even the thermodynamic asymmetry $k$ or shear angle $\theta$ is fixed. The  reconnection rate denotes the amount of the reconnected magnetic flux per unit time by definition. Thus, it is deeply related to the magnetic energy conversion power. In the proposed self-similar MRX model, the magnetic energy conversion power is proportional to the self-similarly expanding MRX system spatial scale. Hence, it is proportional to the time from the onset of the MRX. The reconnection rate is linearly correlated to the coefficient of the proportionality. The extension speed of the MRX system in $x$-direction is estimated by an intermediate value of the projected Alfv\'{e}n speeds in $x$-direction in both sides of the CS. Note that this extension speed is independent of the reconnection rate. The aspect ratio (thickness in $y$-direction)/(length in $x$-direction) of the MRX outflow region decreases as the reconnection rate decreases (Figure \ref{fig:ro_d-00023-60-k2}). Thus, the larger reconnection rate results in the larger amount of the plasma ingress from the side with smaller $V_{Ax}$ to the other side. This infiltrated plasma across the CS is potentially mixable with the plasma in the larger $V_{Ax}$ side, as pointed in Section \ref{sec:CD}. 

Figure \ref{fig:logR*60k1-2-4-8-16-ave} clearly shows also that the reconnection rate sensitively depends on the azimuthal direction $\phi$ of the X-line; even the thermodynamic asymmetry $k$ or the shear angle $\theta$ is fixed. The reconnection rate is maximum at a certain optimized $\phi$ for given $k$ and $\theta$. The case with this optimized $\phi$ should correspond to the fastest growing mode of the spontaneous MRX as a catastrophic global instability under the fixed $k$ and $\theta$ values. This result let us induce an imagination that, in actual three-dimensional situation, the direction of the X-line will be spontaneously tuned to realize the maximum reconnection rate with the optimized $\phi$ (Section 5.2.5 of Nitta \& Kondoh 2021). This will be inspected by our three-dimensional simulation, which is in progress.

\subsection{Some phenomenological arguments}
\label{sec:phenomena}
We discuss the application of the proposed model to the geomagnetospheric phenomena. In previous geomagnetospheric studies, the mixing of the magnetosheath and magnetospheric plasmas associated with MRX is considered to occur around the diffusion region (e.g., Hesse et al. 2016). We here clarified that, in the asymmetric MRX (asymmetry of the thermodynamic quantities or asymmetry of the sheared magnetic configuration), the sheath (smaller $V_{Ax}$) plasma infiltrates the magnetospheric (larger $V_{Ax}$) side plasmoid. The plasmoid expands self-similarly with time; thus, the amount of the sheath plasma ingress increases in proportional to the time from the onset of the MRX (Figure \ref{fig:ro_d-00023-60-k2up}). This implies that a large amount of the sheath plasma can directly infiltrate the magnetospohere. The amount of the sheath plasma that infiltrates the magnetosphere sensitively depends on the parameters indicating asymmetry (the thermodynamic asymmetry parameter $k$ and azimuthal angle $\phi$ of the X-line). These properties will be observationally verified.  

We also discuss the application of the proposed model to the solar phenomena. In the jet phenomena, e.g., the X-ray jet (Yokoyama and Shibata 1996) and anemone jet (Shibata et al. 2007), the asymmetric MRX seems to take place. The asymmetry of these events will be significantly large. As we mentioned in Figure \ref{fig:ro_d-00023-60-k2}a, the resultant plasmoid is considerably elongated and looks like a jet. This property is consistent with previous observations of the solar jets (e.g., Lei et al. 2019). We can predict that 1) the jet will form in the side with larger $V_{Ax}$ of the CS, 2) the jet comprises the plasmas from the side with larger $V_{Ax}$ in the front half region and the plasmas from the side with smaller $V_{Ax}$ in the rear half region (Figure \ref{fig:ro_d-00023-60-k2}). If quantities characterizing each plasmas (temperature, degree of ionization, density, and others) are clearly different, we can observe the predicted double claw structure.

\subsection{About Petschek model as the current standard model}
\label{sec:Petschek}
The Petschek model (Petschek 1964) is the most famous theoretical model of the so-called fast MRX. The reconnection rate of this model is considerably large compared with that of the so-called slow models, e.g., the Sweet--Parker model (Parker 1957, Sweet 1958). The Petschek model was widely accepted as a standard model for solar flares and geomagnetospheric substorms. The most important property of this model is the ``logarithmic dependence of the reconnection rate'' on the magnetic Reynolds number. This actually implies that the reconnection rate of the Petschek model is approximately a universal constant for all the events ($R^* \sim 10^{-2}$). This result was fascinating as a universal model and strongly impressed many researchers; however, it frequently misleads. Many researchers seem to optimistically believe that the reconnection rate of the realistic MRX is of the order of $10^{-2}$ for any situation. We focus on this point below. 

It is not well known that the famous logarithmic dependence of the Petschek model is based on an unrealistic assumption. The model assumes that the thickness of the diffusion region can unlimitedly decrease responding to the inflow speed when the magnetic Reynolds number increases. This is obviously not true because the thickness of the diffusion region must be bounded by a lower limit estimated as similar to the ion Larmor radius. When the diffusion region thickness is sufficiently larger than the ion Larmor radius, the thickness can decrease as the magnetic Reynolds number increases. Thus, the magnetic diffusion speed can hold. When the diffusion region thickness reaches the lower boundary as the magnetic Reynolds number increases, the diffusion region thickness cannot decrease any more. In this state, the magnetic diffusion speed decreases as the magnetic Reynolds number increases. Thus, the reconnection rate decreases as well. The magnetic Reynolds number dependence of this state is no more the logarithmic as predicted by the Petschek model but is a power law ($R^* \propto {R_{em}}^{-1}$, see Nitta 2007). We must realize that the reconnection rate of fast MRX considerably varies in the actual situation.

\subsection{Reliability of the global MHD simulations}
\label{sec:global-siimulation}
MRX plays important roles in the global MHD simulation for, e.g., the dynamo process in the solar convection zone, the magnetorotational instabilities in accretion disks of a lot of kinds of astrophysical objects, and others. In such global simulations, can we solve the MRX with sufficient resolution? If the resolution is insufficient, is the simulation result reliable? 

In our previous simulations (Nitta et al. 2016, Nitta \& Kondoh 2019, 2021), the initial half-thickness $D$ of the CS is assumed to be similar to the ion Larmor radius. This is expected to simulate the situation that the CS is well compressed and is just on the brink of enhancing the anomalous resistivity. Hence, we treat $D$ as almost the lower limit of the CS thickness. 

We adopt simulation meshes as fine as possible in our allowable computational resources. In typical cases, the initial current sheet half-thickness $D$ is resolved by five meshes (mesh size $=0.2 D$). This is almost the lowest resolution for reliable simulation. We use totally 12800 $\times$ 25600 meshes for the simulation box ($2560D \times 5120D$) and simulate for the temporal duration $2400 D/{V_{A0}}$. If we translate it to the typical values in the solar corona, the current sheet half-thickness (i.e., the ion Larmor radius) is $\sim 1D \sim 10^0$ [m] and the Alfv\'{e}n speed is $V_{A0} \sim 2 \times 10^6$ [m s$^{-1}$]. Thus, our simulation box size corresponds to merely $\sim (2 \times 10^3$ [m]) $\times (4 \times 10^3$ [m]) and performing the temporal evolution for barely $\sim 10^{-3}$ [s]. The readers may be surprised by this; however, this is almost the limit of the current super-computing in early 2020s. We should realize that the reliable simulation of the MRX significantly consumes computational resources. 

The question then arises. If we perform the MHD simulation with insufficient resolution, how does the MRX occur? We show here the typical cases of the physically meaningful MRX and unphysical ``numerical MRX.'' Figure \ref{fig:R*-t-eta} shows the temporal variation of the reconnection rate $R^*$ for the symmetric ($k=1$) and shear-less ($\theta=\phi=0^\circ$) cases with $\beta_0=0.2$, $R_{em}=24.5, 50., 100.$, respectively. The resolution is $1D$=5 meshes (mesh size=0.2$D$) in this trial. 

Figure \ref{fig:R*-t-eta} for $R_{em}=24.5$ shows the physically meaningful (regular) temporal evolution of the reconnection rate for fixed resistivity as the reference. The typical regular temporal evolution is that the reconnection rate once increases (the width of the diffusion region decreases) just after the onset of the MRX and then gradually decreases (the diffusion region thickness increases) after the maximum to an asymptotic value of the self-similar phase. 

On the contrary, we can find an abnormal temporal variation of the reconnection rate, as shown in Figure \ref{fig:R*-t-eta}, for $R_{em}=100.$. The reconnection rate suddenly increases at a temporal timing ($200 < t < 400$ in this case). The reconnection rate after $t=400$ is almost 3.3 times larger than that before $t=200$. We note that the reconnection rate in the self-similar phase is inversely proportional to the magnetic Reynolds number (Figure 4 of Nitta 2007; we analytically investigated the fundamental properties for the shear-less symmetric cases ($\theta=\phi=0$, $k=1$)). Thus, the expected asymptotic reconnection rate of this case is $R^* \sim 0.011$ (similar to the reconnection rate at $t=0-200$). The resultant MRX (with reconnection rate $R^* \sim 0.035$) of this simulation is unphysically faster than the realistic MRX. This is of course a fake numerical MRX. 

Figure \ref{fig:bx-y-eta} shows the $B_x$ profiles along the $y$-axis at $t=0, 200$, and $800$ for $R_{em}=24.5, 50., and 100.$. The dots denote grid points of the meshes. Note that we can treat the number of the dots in the transition region, in which the $B_x$ value suddenly changes as a measure of the resolution (the dots are dense and connected to each other, except the transition region, to form an apparent thick line). The CS is well resolved by the simulation meshes in the initial state at $t=0$. When $t=800$, the diffusion region is fairly resolved by five or six meshes for $R_{em}=24.5$, but failed for $R_{em}=100$ (the diffusion region involves only two meshes). The transition region of $B_x$ is very steep, and the diffusion region is not sufficiently resolved by the meshes for $R_{em}=100$.  

Let us focus on the peculiar behavior of $R_{em}=100$. As shown in Figure \ref{fig:R*-t-eta} for $R_{em}=100$, we can observe a drastic increment in the reconnection rate between $t=200-400$. Figure \ref{fig:bx-y-100} shows what happened in this period. The thinning of the diffusion region proceeds from $t=200$, then the thickness reaches almost the minimum value ($\sim 0.2 D$: about one fifth of the initial thickness), which is similar to the mesh size at $t=400$. Remember we supposed that the initial thickness ($\sim 2D$) corresponds to the physically possible minimum scale (similar to the ion Larmor radius). Thus, such excessive thinning of the diffusion region is unfavorable. Unphysical numerical diffusion may dominate in this case. We understand that this is a typical property of the unphysical ``numerical MRX'' in the case of insufficient resolution. 

As shown in this part, the MRX with insufficient resolution often leads to unphysically fast numerical MRX. Such unphysical behavior by the numerical MRX may not be avoidable in the finite mesh method. It is difficult to solve MRX with the small reconnection rate to remain in slow energy conversion and a slowly magnetic topology changing process. The global astrophysical systems comprise, in general,  a lot of elementary physical processes; each process includes a proper time/spatial scale. The MRX process is one of the smallest and fastest elementary process of such systems. We can easily understand that the reliable global long-term simulations in which a lot of elementary processes having extremely different timescales simultaneously proceed are actually impossible to solve. If we correctly solve the MRX process in the global simulation, it forces us to use extraordinary huge computational resources. If the MRX process is unrealistically fast and its timescale is unrealistically short owing to the lack of resolution, such temporal evolution should be significantly different from the actual phenomena, and is far from convincing. In other words, we must carefully and critically evaluate the results of the global numerical simulations in which the MRX process cannot be correctly solved.

\subsection{Self-similar model as subgrid gimmick for the global MHD simulations}
As discussed in Section \ref{sec:global-siimulation}, the result of the global simulations, in which the resolution of the diffusion region is insufficient, may include a lot of rational doubts. Especially, we have to be skeptical about the temporal evolution of the system because MRX as an elementary process may proceed too fast by the numerical (unphysical) MRX. This implies that non-dimensional parameters defined by the ratio of timescales of each elementary process cannot be set correctly in the global simulation. Based on the similarity law, such simulations break down. 

We believe that with our current computational resources, it is difficult (or impossible) to precisely solve up to the MRX level (even in the MHD framework) via global MHD simulations. MRX with insufficient resolution results in unphysically fast energy conversion and magnetic topology changes. These may lead the system to unphysical evolution owing to the numerical MRX. How can we clear this serious problem? 

In the current status, there are probably two ways to solve this problem. The first one is using the adaptive mesh method around the X-line as in the numerical simulation with respect to the stellar forming region (e.g., Winkler, Norman and Newman 1984). If we detect the abnormal thinning of the diffusion region and unphysical numerical diffusion is likely to occur, the simulation code automatically adds sufficient fine meshes around the thinning diffusion region. Then, the code reruns just before the initiation of the abnormal thinning. The throat of this method might be that the diffusion region should be sufficiently resolved during the entire MRX temporal evolution process. Such simulation must be considerably time/resource-consuming. This point is quite different from, e.g., the dynamical process of the stellar-forming region, in which small structures are formed in the later stage of the collapse. The case in which the abnormal diffusion region thinning occurs may come up in the very slowly evolving MRX with, e.g., the high magnetic Reynolds number, the highly asymmetric, or the highly sheared circumstances. In such cases, the X-line may move at a considerable speed compared with the ambient Alfv\'{e}n speed in the direction along the CS (see Nitta 2007 for cases with high magnetic Reynolds number) or perpendicular to the CS (Nitta \& Kondoh 2019 and Nitta \& Kondoh 2021 for asymmetric cases). We must follow the motion of the X-line to create relevant adaptive meshes. This may not be easy to accomplish. 

The second one is inserting MRX virtual effects into the simulation code instead of precisely solving the MRX process in global simulations. A similar procedure is adopted in the global simulation of galaxy formation studies (so-called ``subgrid model'' reproducing the virtual effects of star formation or supernovae; e.g., as classical papers, Katz 1992, Navarro et al. 1995, Mori et al. 1997, as recent paper, Pillepich et al. 2018). It is not realistic to treat small-scale star formations or supernovae in the simulations of the entire galactic scale. We must prepare a relevant gimmick as the subgrid model that virtually represents the physical effects of the star formations and supernovae. This idea will be also effective for our purpose, i.e., we adopt a gimmick reproducing the effects of the MRX process that is embedded in the global MHD simulation code as a subgrid model. This gimmick must perform the magnetic flux transfer and energy conversion by MRX in a relevant timescale without solving the MRX process itself. We do not further discuss this; however, our series of studies on the self-similar MRX model may provide a new step toward this direction and may help contribute to future related studies. 

\begin{acknowledgments}
The original idea of this study had been stimulated by private discussion with Kanya Kusano (Institute for Space-Earth Environment Research (ISEE), Nagoya University). This study is supported by MEXT/JSPS KAKENHI under Grant No. 21K03645. Numerical computations were carried out using Cray XC50 at the Center for Computational Astrophysics, National Astronomical Observatory of Japan; the KDK computer system at RISH, Kyoto University; and the Fujitsu (FX10 / CX400) system at the Information Technology Center, Nagoya University. S.N. would like to especially thank Ippon-Kakou-Kai for motivation.
\end{acknowledgments}

\appendix
\section{Comparison of reconnection rate in our series of studies}
\label{sec:app-A}
We summarize the reconnection rate as the functions of the parameters and compare the results of our previous studies to clarify the effect of the thermodynamic asymmetry and the magnetic shear on the reconnection rate. 

First, we focus on plasma-beta dependence. We frequently encounter low beta cases in astrophysical problems. For example, the typical plasma-beta value in the solar corona and geomagnetosphere is $\sim 10^{-2}$. However, we cannot treat the cases with even plasma-beta $\sim 10^{-1}$ by numerical simulations based on the finite difference method. The current numerical simulation is ineffective for extremely low- or high-beta MRX. The knowledge for the extremely low- or high-beta MRX was first introduced by Nitta (2004, 2006). He adopted a semi-analytical method instead of the numerical simulation. He set a kind of Riemann (shock-tube) problem along the reconnection outflow and inflow toward the diffusion region. By solving this Riemann problem, he succeeded to clarify the MRX in a wide range of plasma-beta (Nitta 2006 demonstrated for $10^{-4}-10^3$, which cannot be treated by numerical simulations). We can extend the range for further low- or high-beta cases if we continue the calculation.

Figure 2 of Nitta (2006) shows the plasma-beta dependence of the reconnection rate of the shear-less symmetric Petschek-type MRX (in which only one X-point is formed). The reconnection rate does not significantly change owing to a considerable variation in plasma-beta. We can observe a trend that the reconnection rate is higher in the low-beta cases than that in the high-beta cases. 

Second, we focus on the magnetic Reynolds number dependence. The magnetic Reynolds number dependence of the reconnection rate is clarified in Nitta (2007) for the shear-less symmetric case. We define the magnetic Reynolds number as in equation (\ref{eq:Rem}) of this paper. Note that the denominator of equation (\ref{eq:Rem}) is the actual magnetic diffusion speed. It is very difficult to control the magnetic diffusion speed as a handling parameter because the diffusion region thickness temporally varies to respond to the variation in the electrical resistivity $\eta$. Thus, we cannot artificially control the magnetic Reynolds number in numerical simulations. 

Nitta (2007) adopted a semi-analytical method (the shock-tube approximation), as described in Nitta (2004, 2006). By solving this Riemann problem, he succeeded to clarify the dependence of the reconnection system to the magnetic Reynolds number in a very wide dynamic range ($10^0 \leq R_{em} \leq 10^3$). Figure 4 of Nitta (2007) shows the magnetic Reynolds number dependence of the reconnection rate. We can observe a typical power-law dependence that the reconnection rate is inversely proportional to the magnetic Reynolds number $R^* \propto {R_{em}}^{-1}$. This series of solutions exhibits a continuous transition from the Petschek-like fast scheme to the Sweet-Parker-like slow scheme as the magnetic Reynolds number increases. 

Third, we focus on the dependence on the asymmetry of thermodynamic quantities. In Nitta et al. (2016) and Nitta \& Kondoh (2019), we investigated the effect of the asymmetry of the magnetic field strength and thermodynamic quantities with regard to the CS in the shear-less situation. The reconnection rate sensitively depends on the thermodynamic asymmetry parameter $k$. The $k$-dependence of the reconnection rate $R^*$ for $\theta=\phi=0^\circ$, $\beta_0=0.2$, and $R_{em}=24.5$ is shown in Figure 13 of Nitta et al. (2016) and Figure 8 of Nitta \& Kondoh (2019). However, as we found a severe error in the data post-process to evaluate the reconnection rate (Note that the simulation itself has no problem; the problem is in the process evaluating the reconnection rate.), we retreat the process and show the correct figure representing the thermodynamic asymmetry $k$-dependence of the reconnection rate $R^*$ in Figure \ref{fig:logR*-logk-0-0} of this paper for the shear-less case $\theta=\phi=0$, $\beta_0=0.2$, $R_{em}=24.5$. We can observe a typical power-law dependence of the reconnection rate $R^*$ on the thermodynamic asymmetry parameter $k$ ($R^* = 0.0393 \ k^{-1.41}$), as reported in Nitta \& Kondoh (2019). The reconnection rate in the symmetric case ($k=1$) is consistent with regard to the prediction $R^* \sim 10^{-2}$ of the Petschek model, but is significantly decreasing depending on the thermodynamic asymmetry. 

\begin{figure}
\vspace{-1.5\baselineskip}
\begin{center}
\includegraphics[width=12cm, angle=0]{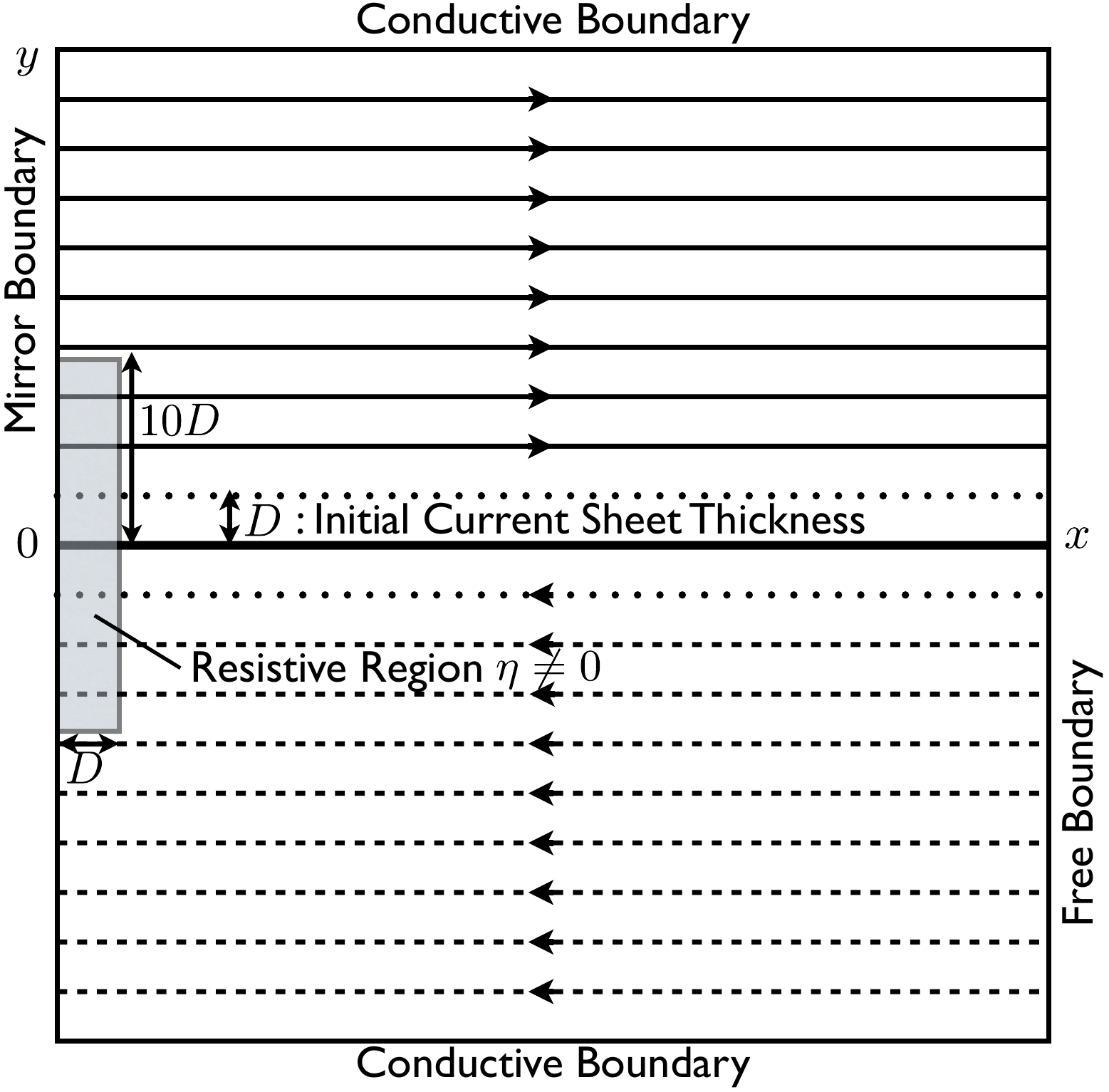}
 \caption{
Schematic picture of the simulation box. The initial CS horizontally lies in the middle of the simulation box. The origin $0$ of the simulation coordinate is the midpoint of the left boundary. The $x$-axis is rightward along the initial CS. The $y$-axis coincides with the left boundary. The $z$-axis is in the foreground direction. In our two-dimensional simulation, we suppose translational symmetry in the $z$-direction. The so-called ``free-boundary'' condition is imposed on the right boundary. The so-called ``conductive wall boundary'' condition is imposed on the top and bottom boundaries. The mirror symmetry is imposed on the $y$-axis (the mirror boundary). The resistive region is indicated by the gray shaded region around the origin. The electrical resistivity $\eta$ is a finite constant of the time and uniform in the resistive region. The resistive region is vertically long and exhibits a rectangular shape to follow a possible vertical motion of the X-point. 
}
\label{fig:box}
\end{center}
\end{figure}

\begin{figure}
\vspace{-1.5\baselineskip}
\begin{center}
        \includegraphics[width=15cm, angle=0]{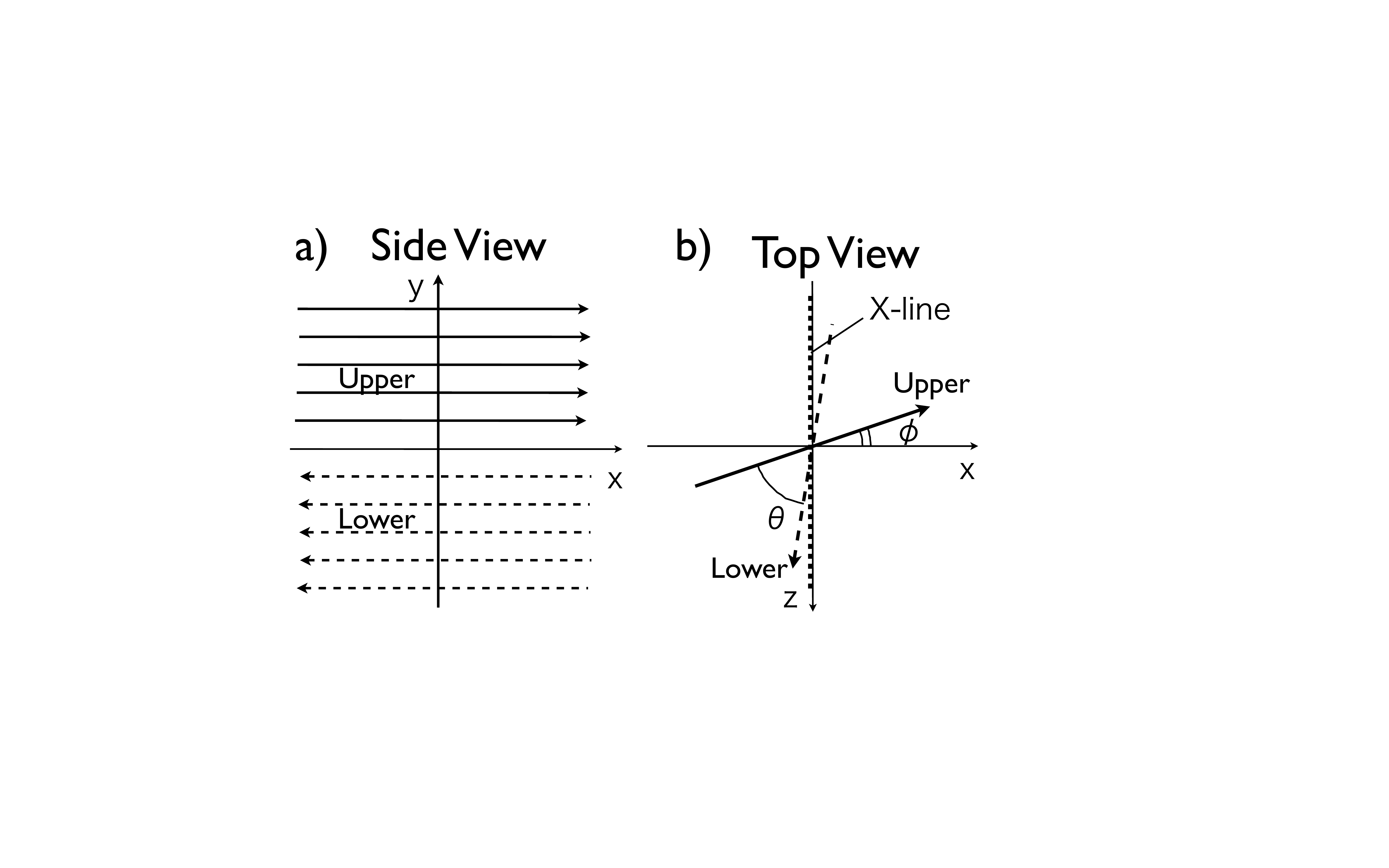}
 \caption{
The definition of the relative angle $\theta$ (the shear parameter) and azimuthal angle $\phi$. The solid and dashed arrows denote the magnetic field lines in the upper and lower region, respectively. 
a) Magnetic configuration observed from the $z$-axis (side view). 
b) Magnetic configuration observed from the $y$-axis (top view). We introduce another azimuthal angle $\phi'$ defined as $\phi' \equiv \phi-\theta/2$. Note that, in our two-dimensional simulations, the reconnection line (the X-line) is always parallel to the $z$-axis. The case $\theta=0^\circ$ denotes the shear-less anti-parallel configuration. The case $\phi'=0^\circ$ denotes the geometrically symmetric configuration with regard to the CS. 
}
\label{fig:concept}
\end{center}
\end{figure}

\begin{figure}
\vspace{-1.5\baselineskip}
\begin{center}
\includegraphics[width=15cm, angle=-90]{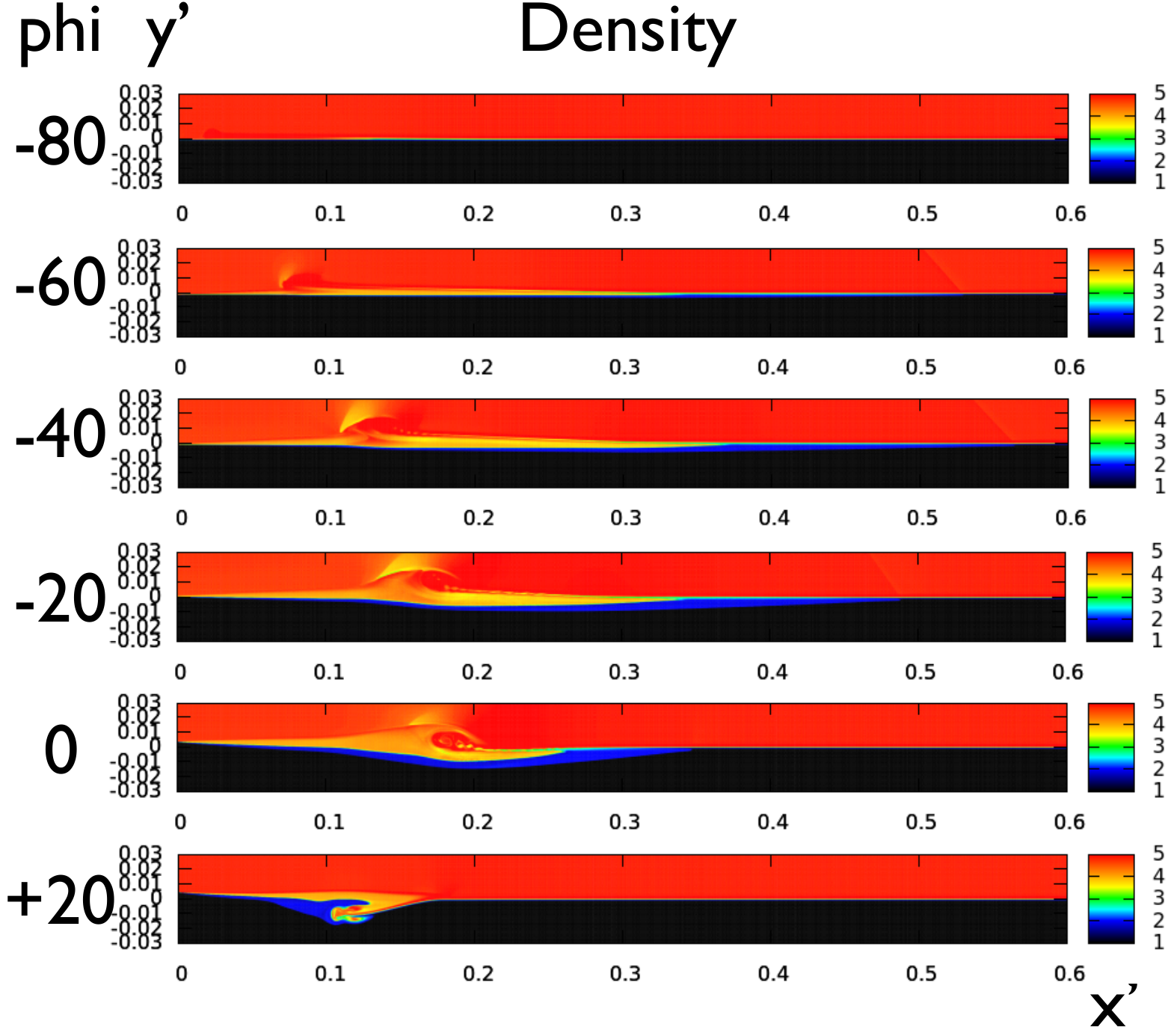}
 \caption{
 Density distribution maps for $\theta=60^\circ, \phi=-80^\circ-+20^\circ$, $k=2$ at $t=2300$. In the panels for $\phi=-80^\circ-0^\circ$, the $x$-direction projected Alfv\'{e}n speed $V_{Ax}$ is larger in the lower half region ($y<0$), whereas in the panel for $\phi=+20^\circ$, $V_{Ax}$ is larger in the upper half region ($y>0$). We can clearly find a double crab-hand claw structure in the lower plasmoid in $\phi=-60^\circ-0^\circ$ where the dense plasma coming from the upper (higher beta) region (e.g., $\phi=-20^\circ$, orange small claw in $\sim 0.17 < x' < 0.34$) infiltrates the lower plasmoid (for $\phi=-20^\circ$, blue large claw $\sim 0.15 < x' < 0.48$) to form a CD (the clear boundary between the forwarding blue plasma and infiltrated orange plasma). The double claw structure is also formed in $\phi=-80^\circ$. However, it is hard to observe in this figure because the reconnection rate is so small that the outflow structure is very thin. We can also find a similar double crab-hand claw structure in the upper plasmoid in $\phi=+20^\circ$. The tenuous plasma coming from the lower (lower beta) region (blue small claw in $\sim 0.1 < x' < 0.12$) infiltrates the upper plasmoid (orange large claw $\sim 0.1 < x' < 0.16$) to form a CD. These double crab-hand claw structures may result in the plasma ingress and mixing across the CS. The yellow region ($0.12<x'<0.16$, $y'>0$) in $\phi=-20^\circ$ above the upper plasmoid is the depressurized region in the total pressure, in which the fast-mode expansion dominates. The sharp rear (left) end edge of this yellow region is the slow shock like the so-called ``normal shock'' formed on the wing of a transonic aircraft. The same structure is found in $\phi=-60^\circ-0^\circ$. 
}
\label{fig:ro_d-00023-60-k2}
\end{center}
\end{figure}

\begin{figure}
\vspace{-1.5\baselineskip}
\begin{center}
        \includegraphics[width=15cm, angle=0]{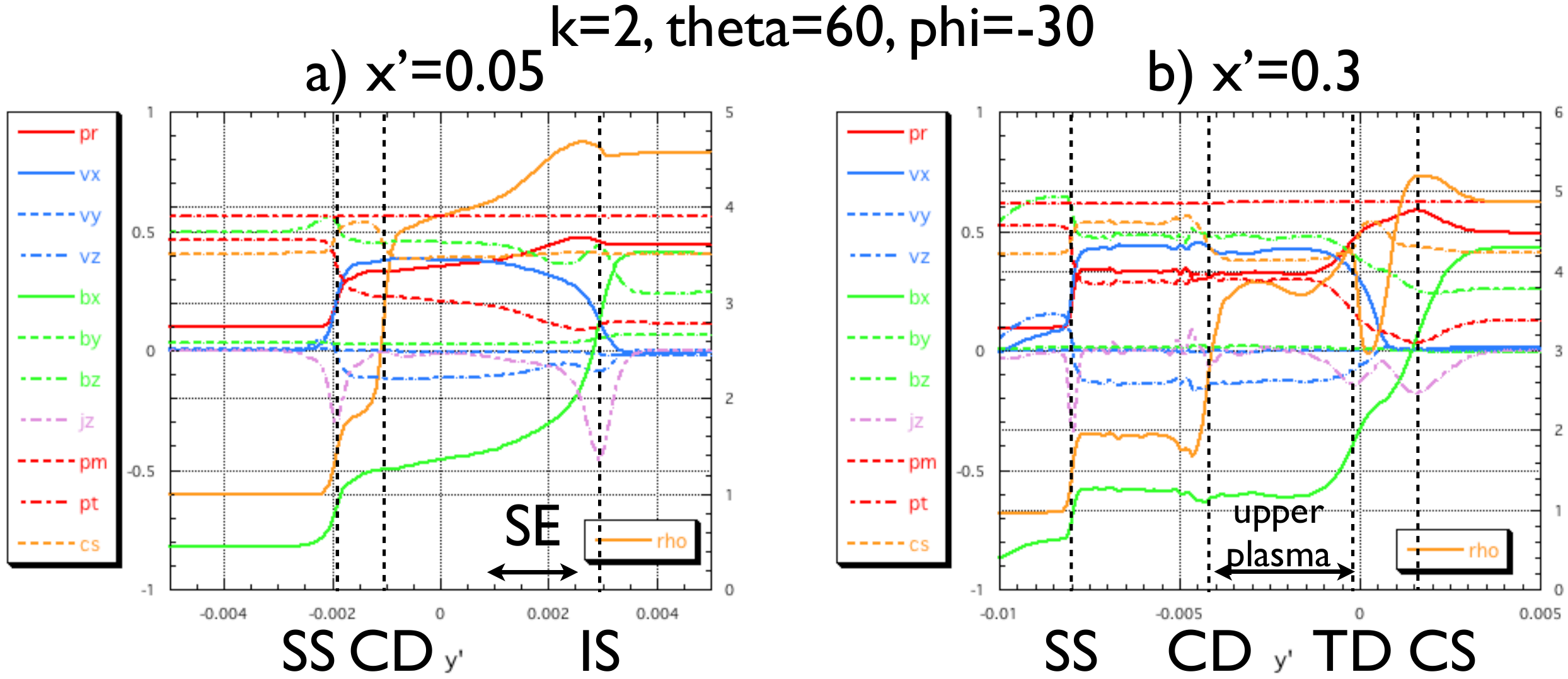}
 \caption{
Profile of the physical quantities at a) $x'=0.05$ (the jet) and b) $x'=0.3$ (the plasmoid) for $k=2$, $\theta=60^\circ$, and $\phi=-30^\circ$. The horizontal axis denotes $y'$. The jet structure (see a)) is qualitatively the same as that with the shear-less asymmetric case shown in Figure 3 of Nitta \& Kondoh (2019). The jet is driven by the SS and combination of the intermediate shock (IS) followed by the slow-mode expansion fan (SE). The jet comprises the two-layered structure separated by the contact discontinuity (CD) between the low-density lower plasma the high-density upper plasma. The upper plasmas infiltrate the lower longer plasmoid (see b), see also Figure \ref{fig:ro_d-00023-60-k2}a)). The infiltrated plasma is surrounded by the tangential discontinuity (TD) and the CD. This is also the same as that with the shear-less asymmetric case of Nitta \& Kondoh (2019). 
}
\label{fig:yprof-60-k2-30}
\end{center}
\end{figure}

\begin{figure}
\vspace{-1.5\baselineskip}
\begin{center}
        \includegraphics[width=15cm, angle=0]{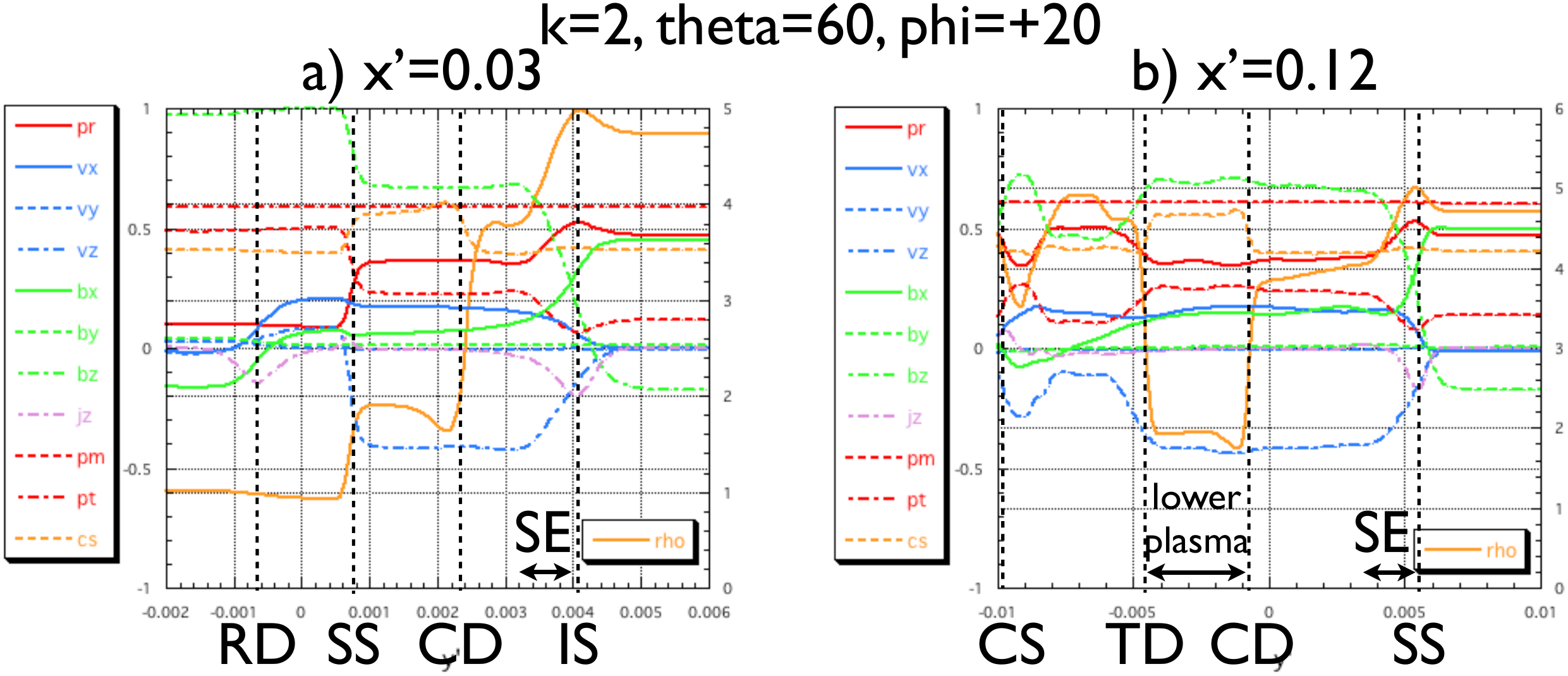}
 \caption{
Profile of the physical quantities in the jet and upper longer plasmoid at a) $x'=0.03$ (the jet) and b) $x'=0.12$ (the plasmoid) for $k=2$, $\theta=60^\circ$, and $\phi=+20^\circ$. The horizontal axis denotes $y'$. The magnetic field strength is larger in the lower region of the current sheet (CS); however, the $x$-component $B_x$ of the magnetic field in the lower region is very small. Thus, the $x$-direction projected Alfv\'{e}n speed in the upper region is larger. This is a proper case of the sheared asymmetric MRX. The jet is driven by the rotational discontinuity (RD) and combination of the intermediate shock (IS) followed by the slow-mode expansion fan (SE). We can observe that the low-density lower plasmas infiltrate the upper plasmoid (see b), see also Figure \ref{fig:ro_d-00023-60-k2}b). The infiltrated plasma is bounded by the TD and CD. 
}
\label{fig:yprof-60-k2+20}
\end{center}
\end{figure}

\begin{figure}[htbp]
\begin{minipage}[b]{0.45\linewidth}
\includegraphics[width=7cm, angle=0]{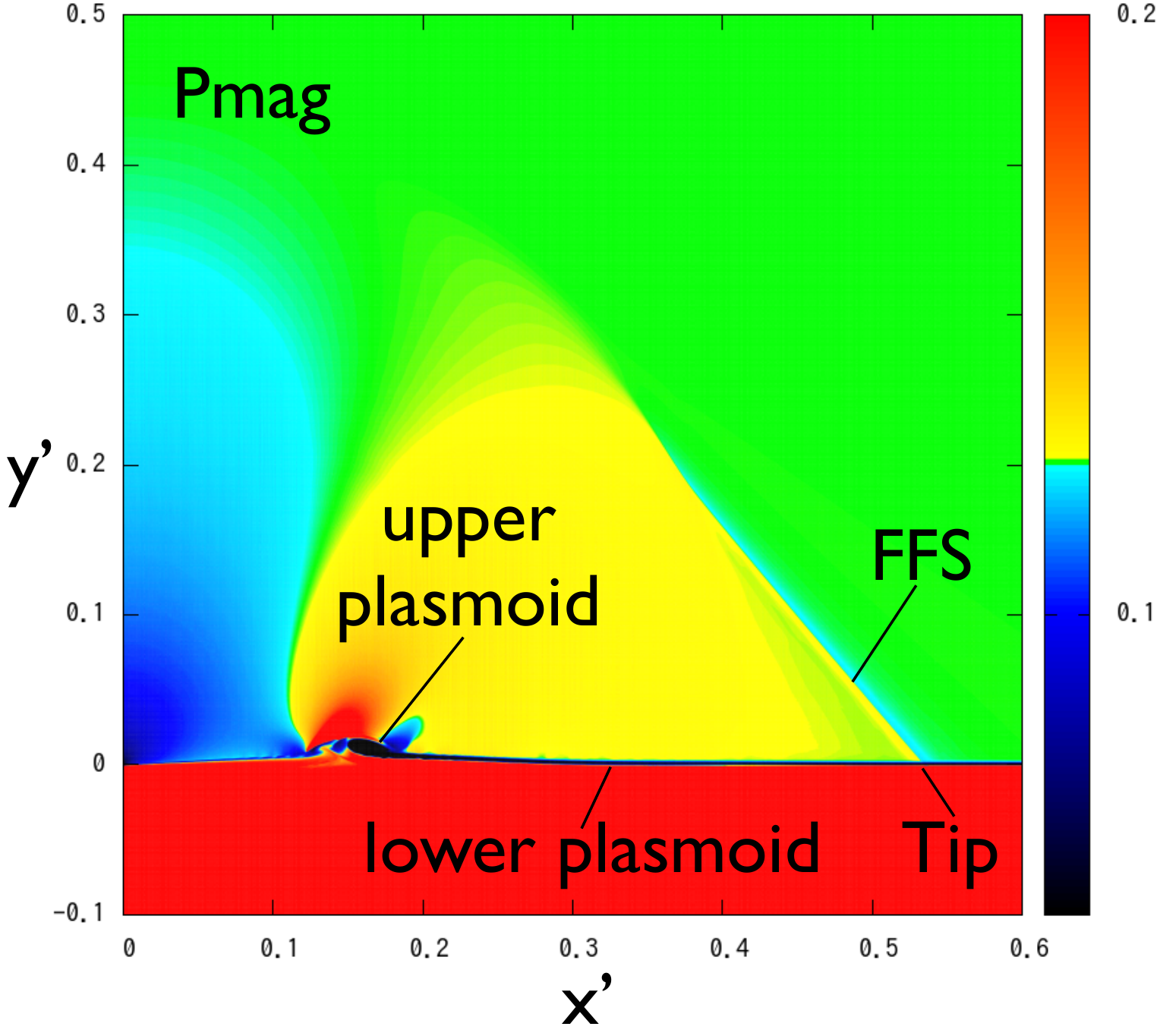}
\end{minipage}
\begin{minipage}[b]{0.45\linewidth}
\includegraphics[width=8cm, angle=0]{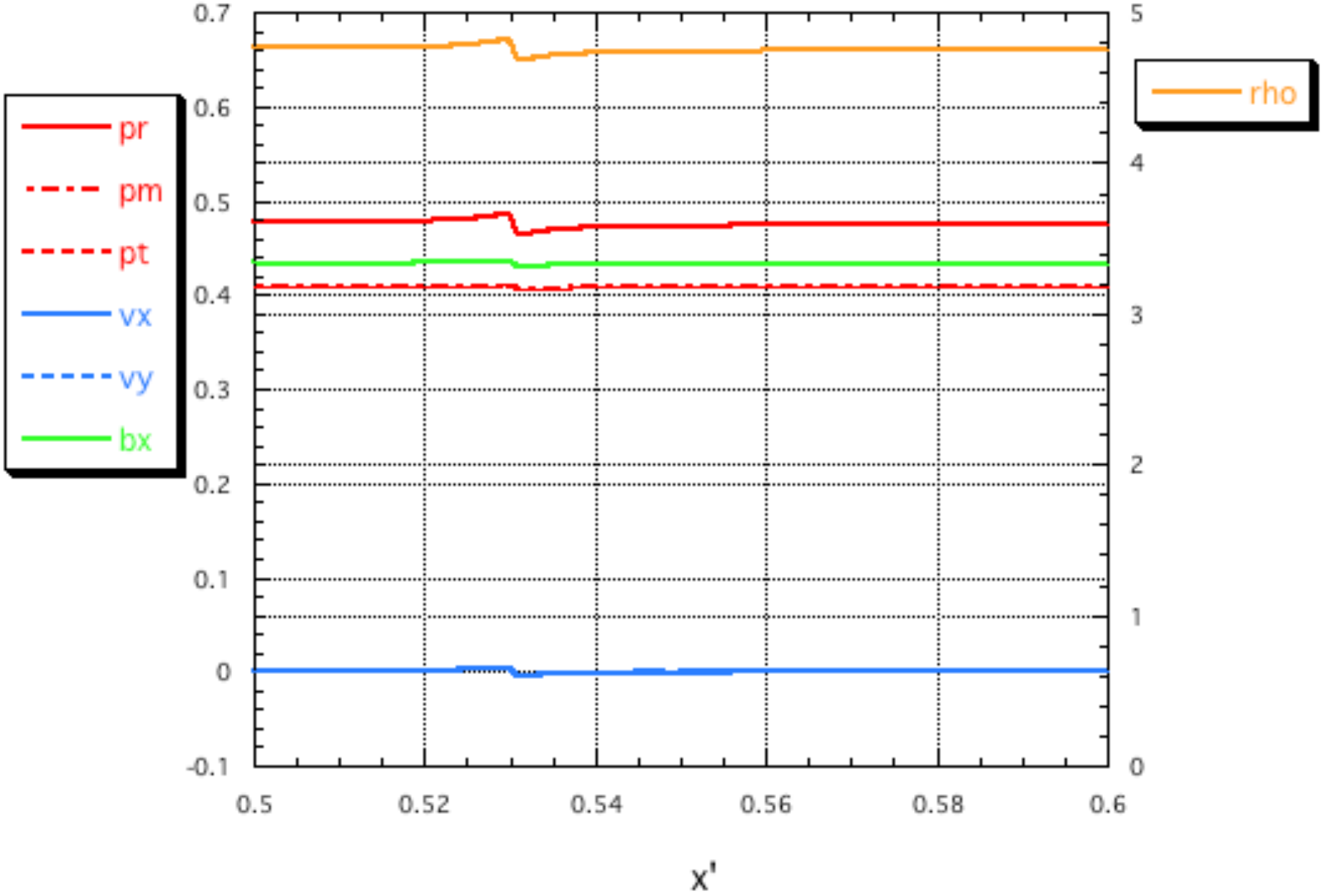}
\end{minipage}
\caption{
a) Magnetic pressure $P_{mag}$ distribution map for $\theta=60^\circ, \phi=-30^\circ$, and $k=2$ at $t=2300$. The horizontal and vertical axes denote $x'$ and $y'$, respectively. The system is approximately in the self-similar phase. We can observe a clear FFS in front of the upper plasmoid. The FFS is produced at the lower plasmoid tip. Its elongating speed ($\sim 0.53 V_{A0}$) exceeds the fast-mode speed ($\sim 0.41 V_{A0}$) of the upper half region ($y>0$). Thus, the fast-mode Mach number of the FFS $\sim 1.3$ (the largest value near the lower plasmoid tip. The fast Mach number decreases along the FFS from the contact point with the CS to upward.). b) Profiles of physical quantities along $y'=0.005$ for the same case of a). A clear but weak jump of the quantities is formed at $x' \sim 0.53$. This is the FFS. The FFS of this sheared case is rather weaker than the FFS of the shear-less case described in Nitta et al. (2016) (the fast-mode Mach number is $\sim 1.6$). d
}
\label{fig:pm_d-00023-60-k2-30}
\end{figure}

\begin{figure}
\vspace{-1.5\baselineskip}
\begin{center}
\includegraphics[width=15cm, angle=0]{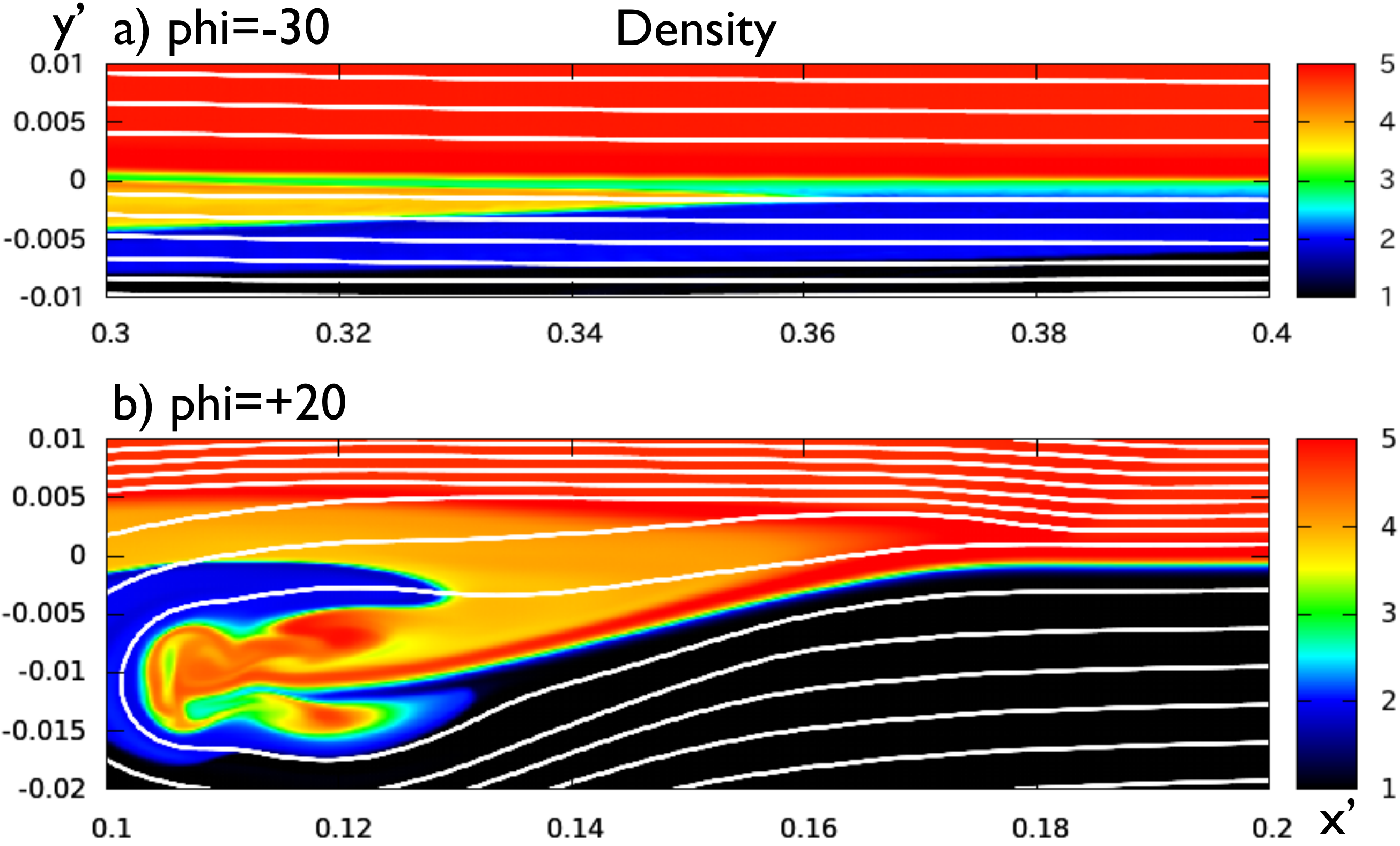}
\caption{
Close-up view of Figure \ref{fig:ro_d-00023-60-k2}: the density distribution map with magnetic field lines (white curves) for a) $\theta=60^\circ, \phi=-30^\circ$, $k=2$ and b) $\theta=60^\circ, \phi=+20^\circ$, $k=2$. The magnetic field lines thread the CD (the boundary between orange and blue in a) and b)). In a), the dense upper plasma (orange) and tenuous lower plasma (blue) coexist on the same magnetic field lines in the lower crab-hand plasmoid. Also in b), the dense upper plasme (orange) and tenuous lower plasma (blue) coexist on the same magnetic field lines in the upper crab-hand plasmoid. These blue and orange components are separated by the CS initially, and then MRX proceeds; these components touch each other on the same magnetic field lines. This may lead plasma mixing through a possible breakup of the plasmoid structure. 
}
\label{fig:ro_d-00023-60-k2up}
\end{center}
\end{figure}

\begin{figure}
\vspace{-1.5\baselineskip}
\begin{center}
\includegraphics[width=12cm, angle=0]{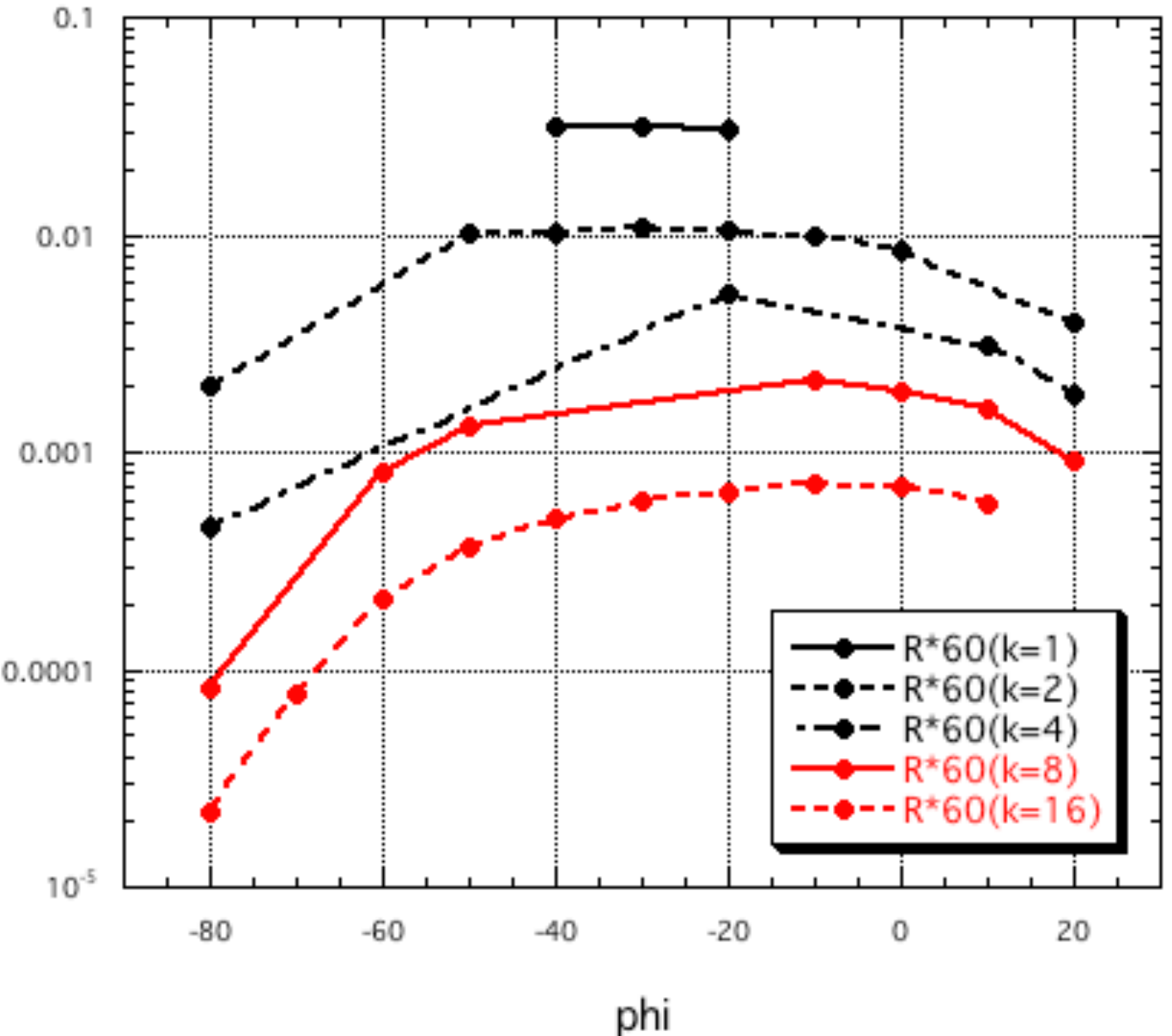}
 \caption{
Azimuthal direction $\phi$-dependence of the reconnection rate $R^*$ with $\theta=60^\circ$ for the thermodynamic asymmetry parameter $k=1, 2, 4, 8, 16$. The reconnection rate drastically decreases as the thermodynamic asymmetry $k$ increases (a power-law dependence as shown in Figure \ref{fig:logR*-logk-0-0}). The asymmetry of the thermodynamic quantities strongly slows MRX. We note that the decrement in the left side ($\phi \sim -90^\circ$) is remarkable compared to the right side ($\phi \sim +30^\circ$) for larger $k$. 
}
\label{fig:logR*60k1-2-4-8-16-ave}
\end{center}
\end{figure}

\begin{figure}[htbp]
  \begin{minipage}[b]{0.45\linewidth}
    \includegraphics[width=7cm, angle=0]{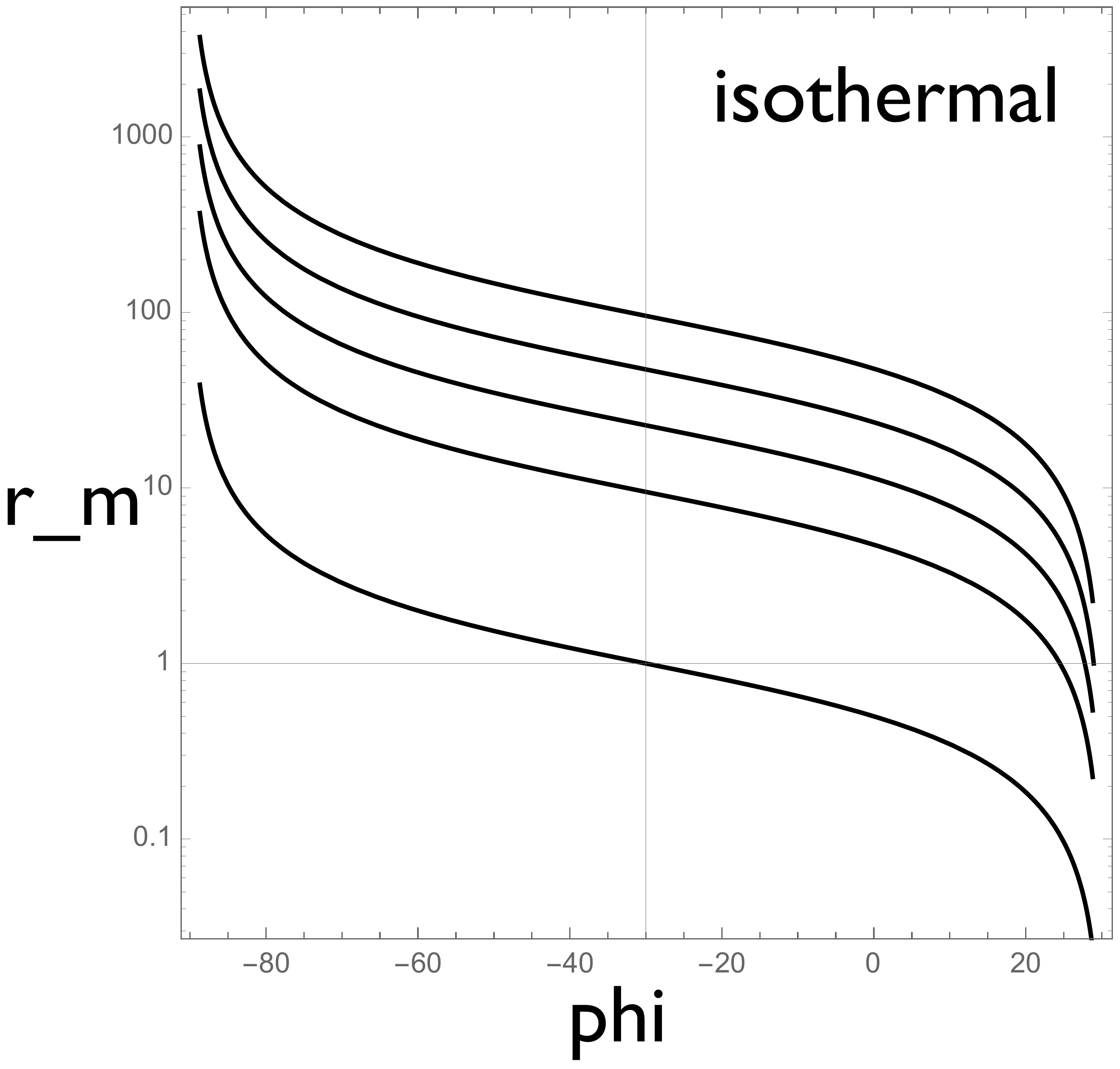}
  \end{minipage}
  \begin{minipage}[b]{0.45\linewidth}
    \includegraphics[width=7cm, angle=0]{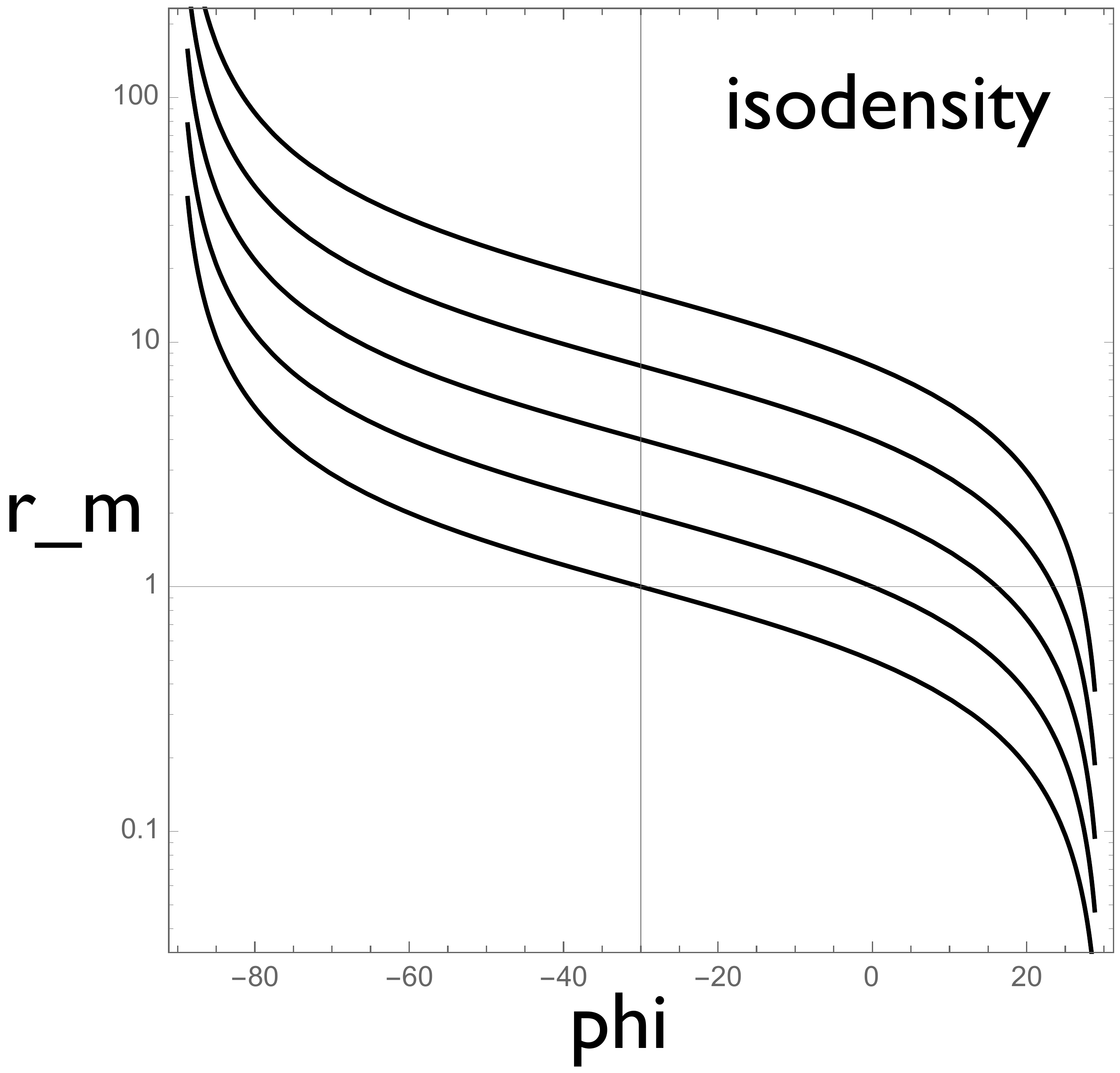}
  \end{minipage}
  \caption{
Azimuthal direction $\phi$-dependence of the mass flux ratio $r_m$ of the upper plasma component to the lower one in the outflow for the a) isothermal and for b) isodensity initial equilibrium cases for the thermodynamic asymmetry parameter $k=1, 2, 4, 8, 16$ (from lower to upper curves). The fraction of the upper component dominates the lower one for larger $k$ and smaller $\phi$ in the outflow in both cases. Only for very large $\phi \ (\sim +30^\circ)$, the lower component dominates. 
}
\label{fig:rm-isothermal}
\end{figure}

\begin{figure}
\vspace{-1.5\baselineskip}
\begin{center}
\includegraphics[width=12cm, angle=0]{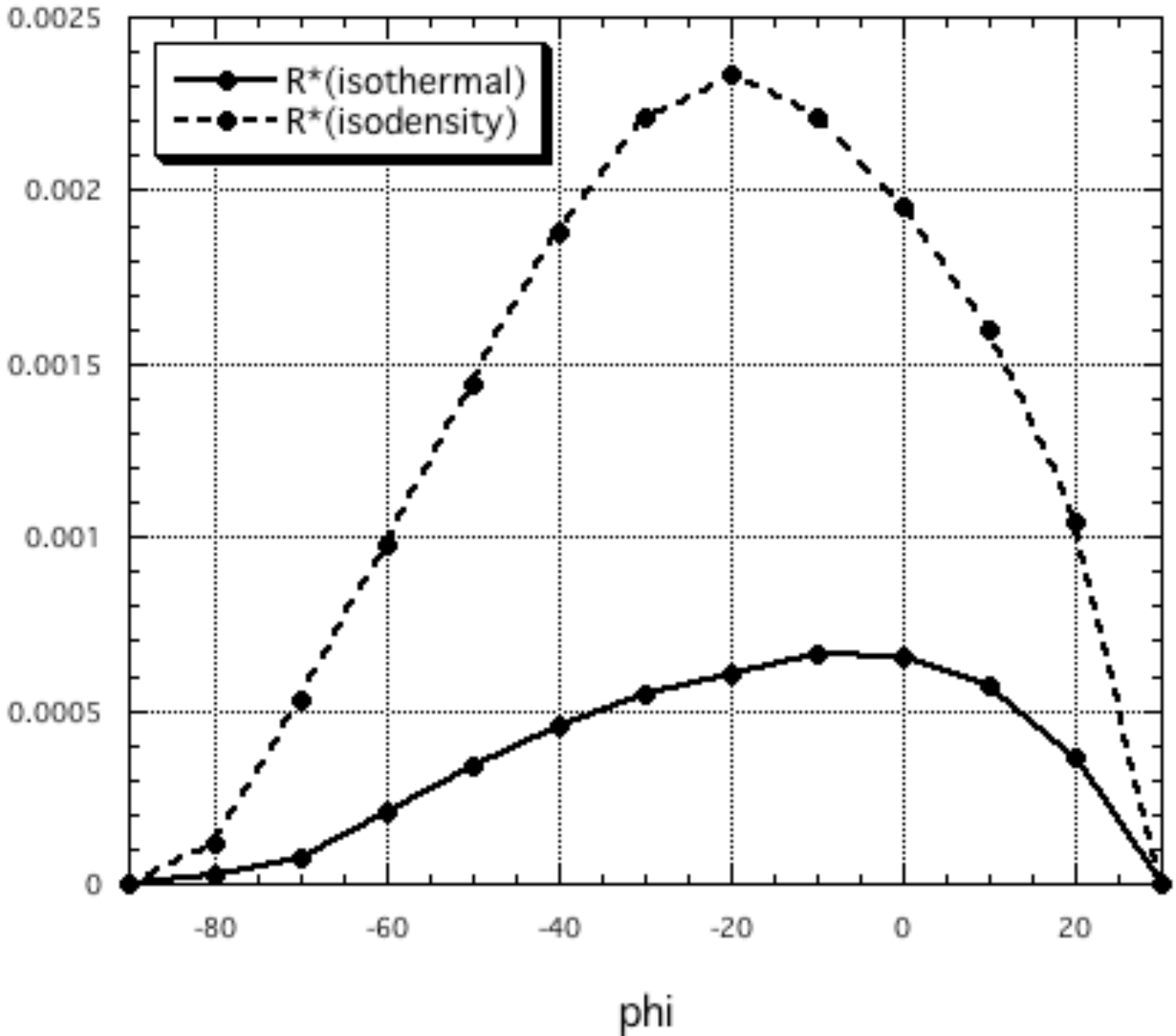}
 \caption{
Azimuthal direction $\phi$-dependence of the reconnection rate $R^*$ of the isothermal initial equilibrium case (solid curve) and the isodensity initial equilibrium case (dashed curve) for the thermodynamic asymmetry parameter $k=16$. 
}
\label{fig:R*-phi-60-k16-isod_isot}
\end{center}
\end{figure}

\begin{figure}
\vspace{-1.5\baselineskip}
\begin{center}
\includegraphics[width=12cm, angle=0]{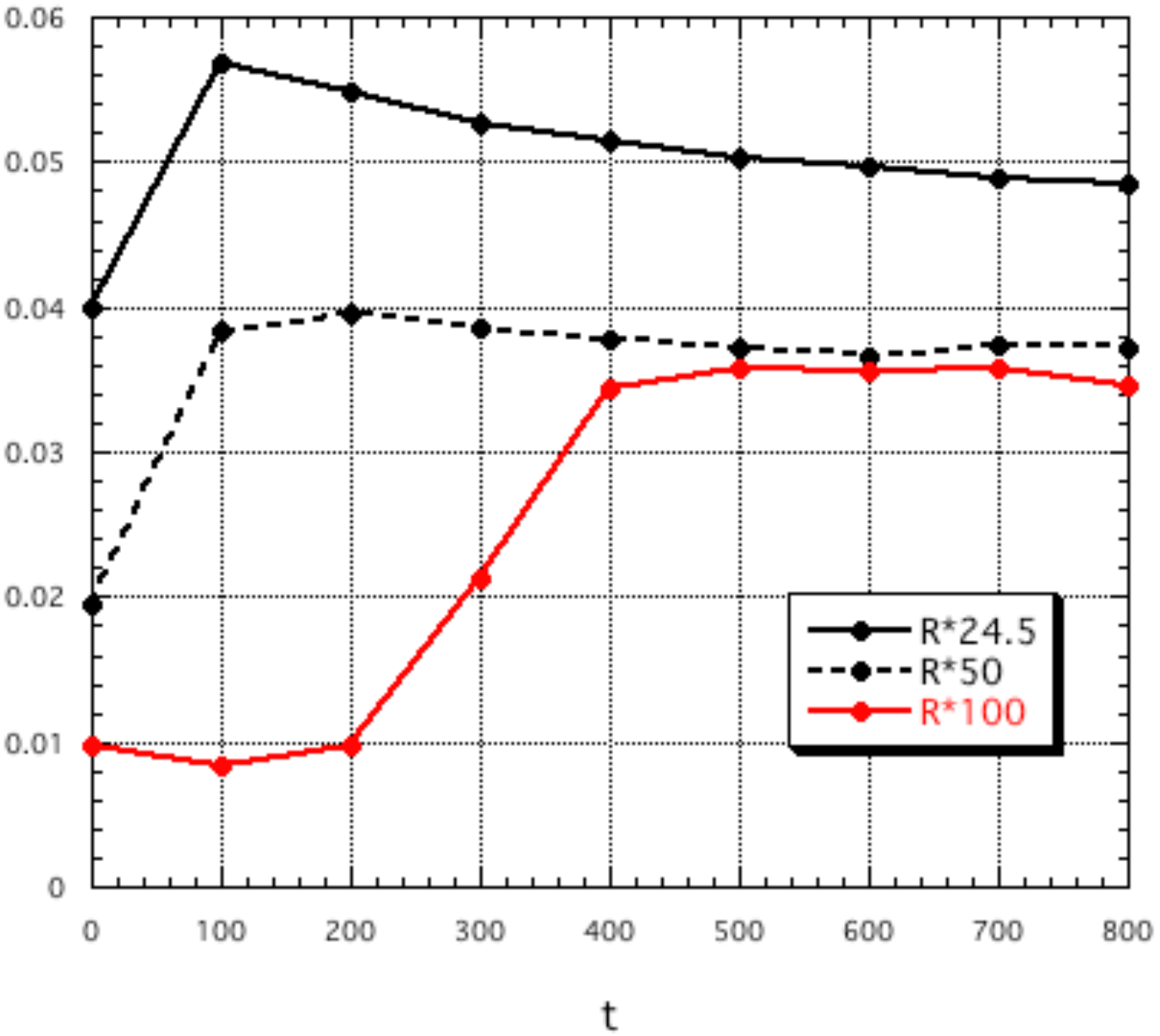}
\caption{
Temporal variation in the reconnection rate $R^*$ for the magnetic Reynolds number $R_{em}=$24.5 (black solid curve), 50 (black dashed curve), 100 (red solid curve) of the shear-less ($\theta=\phi=0$) symmetric ($k=1$) case. The resolution is common to every case (the mesh size is $0.2 D$). The temporal variation of the case $R_{em}=24.5$ is regular for the fixed electrical resistivity models. It is clarified that the asymptotic reconnection rate is approximately inversely proportional to the magnetic Reynolds number $R_{em}$ for the shear-less symmetric case (see Nitta 2007). The temporal variation in $R_{em}=50$ is fairly consistent to this result; however, in $R_{em}=100$, it is obviously extraordinary. The reconnection rate $R^*$ is too large and MRX is unphysically fast. This is the typical behavior of the fake ``numerical MRX''. 
}
\label{fig:R*-t-eta}
\end{center}
\end{figure}

\begin{figure}
\vspace{-1.5\baselineskip}
\begin{center}
\includegraphics[width=15cm, angle=0]{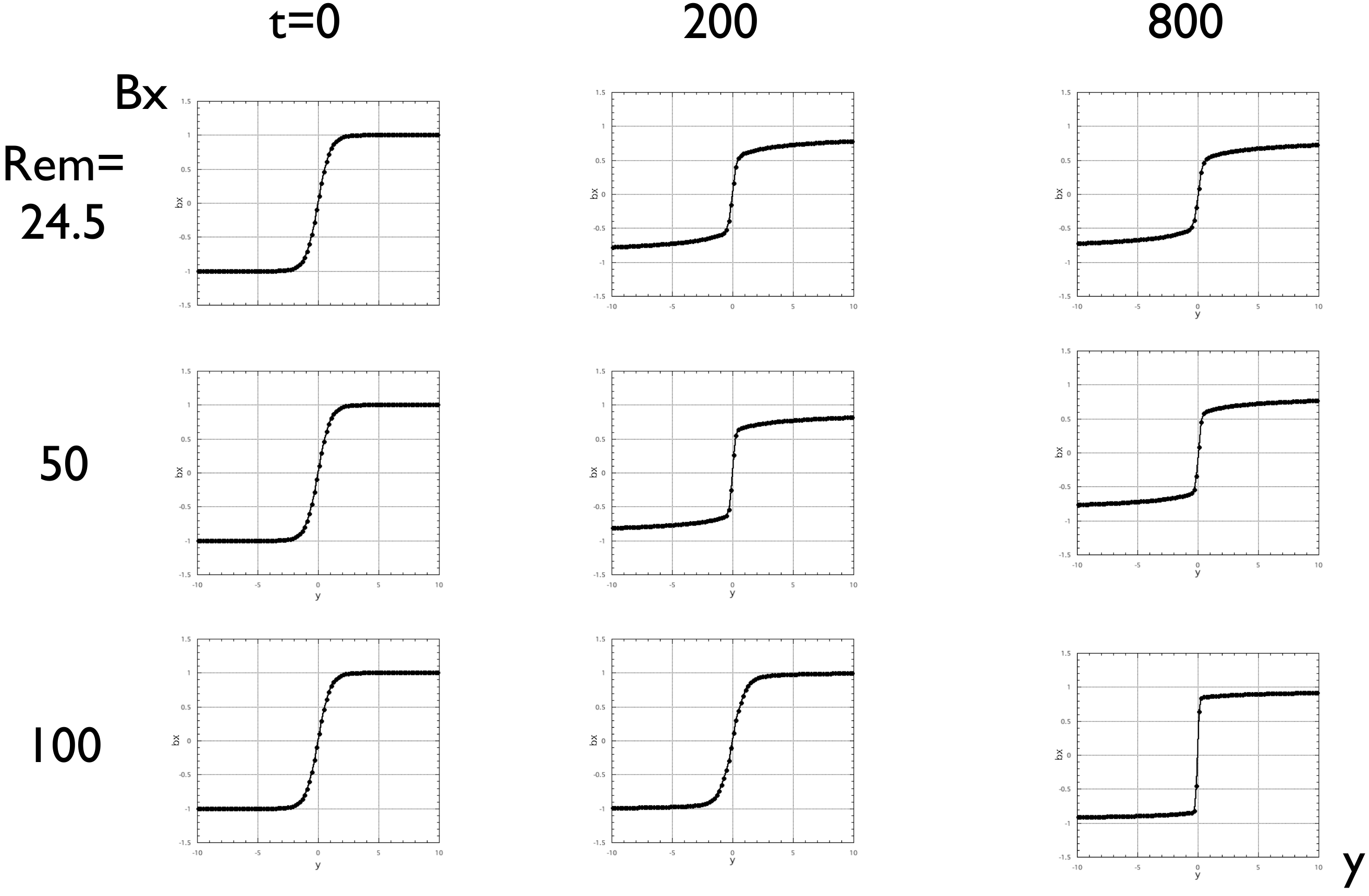}
\caption{
Temporal variation in the $B_x$ profile along the $y$-axis for the magnetic Reynolds number $R_{em}=$24.5 (upper panels), 50 (middle panels), 100 (bottom panels) of the shear-less ($\theta=\phi=0$) symmetric ($k=1$) case for $t=0, 200, 800$ from left to right columns. Dots on the curve denote the grid points of the simulation mesh. These figures show the temporal variation of the diffusion region thickness that is indicated by the width of the transition region of $B_x$, which can be roughly estimated by the number of dots in the transition region. We can find that the transition region is collapsed and is not resolved by meshes in $R_{em}=100$ at $t=800$. It may result in unphysical numerical magnetic diffusion. Such numerical diffusion leads to unphysically fast fake MRX. 
}
\label{fig:bx-y-eta}
\end{center}
\end{figure}

\begin{figure}
\vspace{-1.5\baselineskip}
\begin{center}
\includegraphics[width=15cm, angle=0]{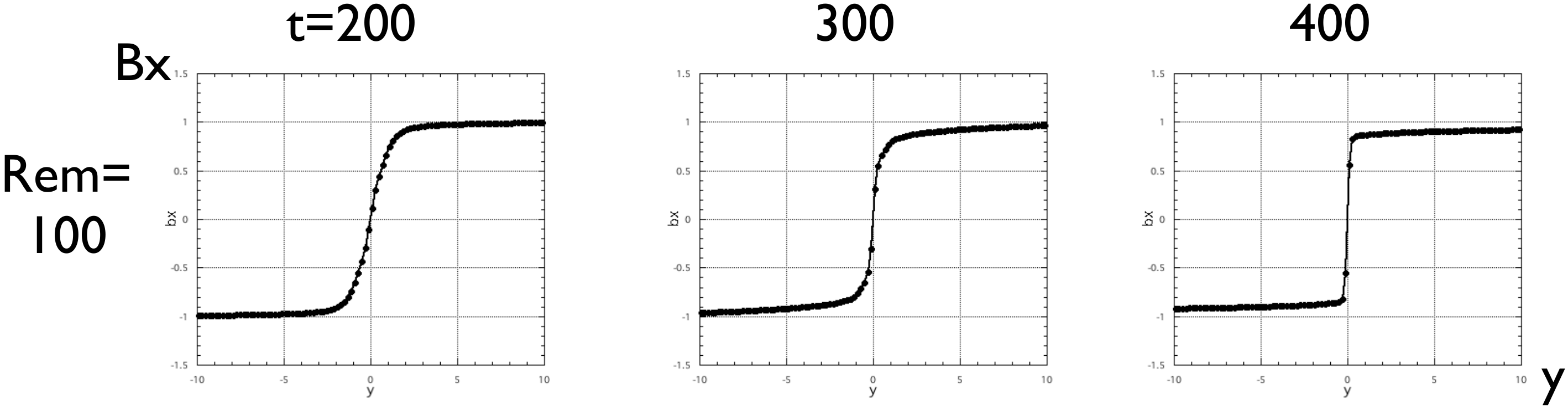}
\caption{
Temporal variation in the $B_x$ profile along the $y$-axis for the magnetic Reynolds number $R_{em}=100$ of the shear-less ($\theta=\phi=0$) symmetric ($k=1$) case for $t=200, 300, 400$ from left to right columns. This figure reveals the process of the unphysical collapse of the diffusion region. The MRX with high magnetic Reynolds number is, in general, hard to solve with the finite difference method. 
\vspace{10cm}
}
\label{fig:bx-y-100}
\end{center}
\end{figure}

\begin{figure}
\vspace{-1.5\baselineskip}
\begin{center}
\includegraphics[width=12cm, angle=0]{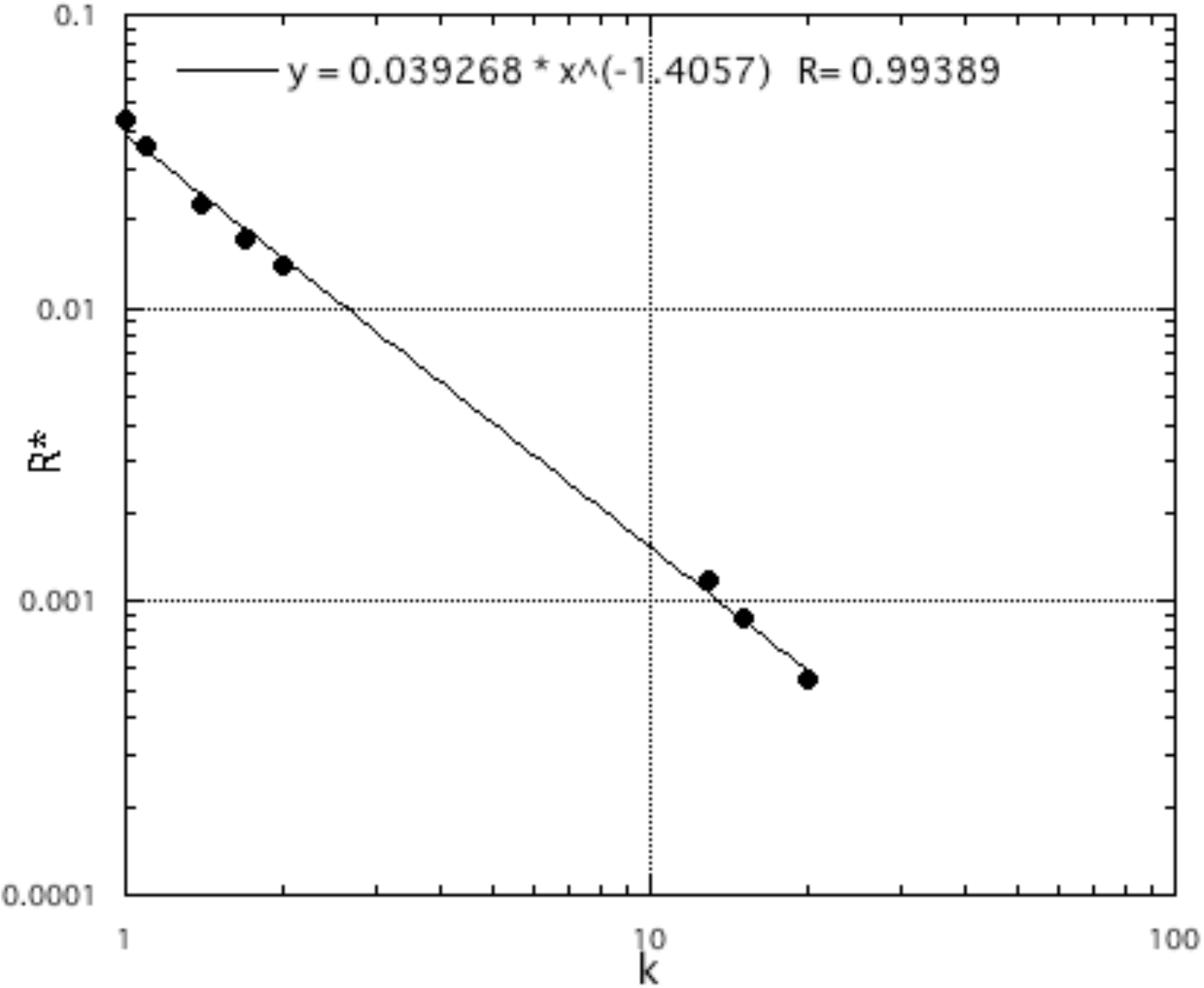}
\caption{
The thermodynamic asymmetry parameter $k$-dependence of the reconnection rate $R^*$ of the shear-less case ($\theta=\phi=0$) for $k=1, 1.1, 1.4, 1.7, 2, 13, 15, 20$. This figure is corrected version of Figure 8a of Nitta \& Kondoh (2019), in which a serious error in the data analysis procedure of the poset process was found. The qualitative properties hold, but the quantitative behavior as the power-law dependence is corrected. 
}
\label{fig:logR*-logk-0-0}
\end{center}
\end{figure}

Fourth, we focus on the magnetic shear dependence. The reconnection rate also depends on the magnetic shear. The magnetic shear is characterized by two parameters: the magnetic shear angle $\theta$ and azimuthal direction $\phi$ of the X-line (Figure \ref{fig:box}). The dependence of the reconnection rate $R^*$ on these parameters has been systematically investigated by Nitta \& Kondoh (2021). We refer to Figures 10 (dependence on $\theta$) and 11 (dependence on $\phi$) of Nitta \& Kondoh (2021). 



\newpage

\newpage

\newpage

\newpage

\newpage

\newpage

\newpage

\newpage

\newpage

\newpage

\newpage

\newpage

\newpage

\newpage

\end{document}